\documentclass[12pt,a4paper]{article}
\usepackage[font=small,skip=4pt]{caption}
\usepackage{placeins}

\usepackage[numbers,sort&compress]{natbib}
\usepackage{url,hyperref,microtype,subcaption,graphicx,booktabs,float,multirow,amsmath}
\usepackage[onehalfspacing]{setspace}
\usepackage{xcolor}
\usepackage{amssymb}
\def\Authors{E. Pace\,$^{1,*}$, L. De~Paolis\,$^{1}$, G. Felici\,$^{1}$, I. Vaglini\,$^{2}$,\\
M. Sandrini\,$^{3}$, M. Pettini\,$^{4}$, A. Gemignani\,$^{2}$ and M. Benfatto\,$^{1,*}$}

\def\Address{$^{1}$Laboratori Nazionali di Frascati, Istituto Nazionale di Fisica Nucleare, Via E. Fermi 54, 00044, Frascati, Italy.\\
$^{2}$Department of Surgical, Medical and Molecular Pathology, Critical and Care Medicine, University of Pisa, Via Savi 10, 56126 Pisa, Italy.\\
$^{3}$Istituto Tecnico ``Morigia-Perdisa'', Via G- Marconi 6, 48124 Ravenna, Italy.\\
$^{4}$Aix-Marseille Universit\'{e}, Centre de Physique Th\'{e}orique (CPT), CNRS UMR 7332, 13288 Marseille, France.}

\def\corrAuthor{Maurizio Benfatto, Elisabetta Pace}
\def\corrEmail{Maurizio.Benfatto@lnf.infn.it, Elisabetta.Pace@lnf.infn.it}

\begin{document}

\title{Biophoton Emission from Palm during Meditation: A Multi-Method Complexity Analysis}

\author{\Authors}

\date{}

\maketitle

{\small\noindent\Address\\[0.5em]
\noindent$^{*}$\textit{Corresponding authors:} \corrAuthor\ (\corrEmail)}
\vspace{12pt}

\begin{abstract}
Biophotons are ultra-weak photon emissions in the visible spectrum produced by
living organisms. While extensively studied in plants, germinating seeds, and cell
cultures, in vivo measurements on the human body remain relatively rare, and no
systematic multi-method complexity analysis of human ultraweak photon emission
(UPE) under controlled physiological modulation has been reported to date.
We address this gap by developing and applying a comprehensive analytical framework
to UPE measurements from the right palm of a human subject. Three independent
sessions were conducted on different days, each comprising four consecutive
15-minute phases: a Dark reference, a pre-meditation resting state (Pre), a
structured meditation phase based on the Sama Vritti box-breathing protocol, and
a post-meditation recovery (Post).

The photon count series are analysed with four complementary methods targeting
distinct aspects of the emission dynamics: distributional statistics (Fano factor,
skewness, tail Expected Shortfall); multiscale Fano factor and Allan deviation;
stripe-filtered Diffusion Entropy Analysis (DEA); and R\'{e}nyi entropy with a
Time Reversal test. A central finding of this methodological study is that the
different methods have complementary sensitivities, jointly converging on a
coherent picture: a systematic reduction of emission intermittency during the
meditative phase, consistently detected across all three sessions. The
stripe-filtered DEA places the biological emission in the non-ergodic renewal
regime ($\delta > 0.5$, $\mu \approx 2.5$--$2.9$), consistent with previous
findings on cell cultures and germinating seeds, with a Pre$\to$Meditation
decrease of $\delta$ approaching overall significance (Stouffer $p \approx
0.055$). The R\'{e}nyi analysis reveals two distinct effects: a reduction of
marginal amplitude burstiness ($T_{\rm dir}$) and a simultaneous increase of
sequential pattern structure ($T_{\rm seq}$), the latter interpreted as
entrainment of the emission dynamics to the Sama Vritti breathing rhythm.
These findings are consistent with the dynamical transitions reported in cardiac
complexity during meditation by Tuladhar et al.\ via DEA of heart rate
variability, and with the EEG reorganization during Sama Vritti breathing
reported by Zaccaro et al.\ in a cohort of experienced meditators, suggesting
a coordinated multi-channel physiological response.
The results establish a proof-of-concept framework for the complexity
analysis of human UPE under physiological modulation and identify the most
sensitive analytical observables for future multi-subject studies.
\end{abstract}


\vspace{12pt}

\section{Introduction}

Biophotons, or ultraweak photon emissions (UPE), are spontaneous electromagnetic radiations in the visible spectral range emitted by all living organisms. Their origin is linked to oxidative metabolic processes, primarily involving reactive oxygen species, lipid peroxidation cascades, and the radiative decay of electronically excited molecular intermediates \citep{Popp1, Wijk, CifraPospisil2014}. Two main, not mutually exclusive hypotheses account for their generation: stochastic radiative decay of metabolically excited molecules and the production of coherent electromagnetic fields by biochemical processes \citep{Popp1, Wijk, Slawinski2, Slawinski}. Experimental evidence indicates that cellular stress enhances biophoton emission, supporting both scenarios \citep{Slawinski2, Slawinski}. Interest in UPE has expanded to encompass a wide range of biological systems , from plants and germinating seeds to animal tissues, cell cultures and the human body, with implications for toxicology, health monitoring, and cancer research \citep{Colli1, Colli2, Gallep, Beloussov, Benfatto1, DePaolis1, DePaolis2, Li, Salari1, Salari2, Tsuchida, Tessaro, Popp2}. Moreover, in this context, Nevoit and colleagues' \citep{Nevoit2025} attempt to demonstrate that UPE underlies cellular communication in human body and brain is of particular significance. 

Measurements from the human body have established that UPE intensity is not spatially uniform: the hands and forehead emit at substantially higher rates , typically 10--100 counts per second , than the trunk or lower limbs \citep{van2007spatial,  edwards1989light, van2005multi, van2006anatomic}. The palm of the hand, in particular, has emerged as an informative site for non-invasive physiological monitoring. Systematic studies have documented circadian modulation of emission over 24-hour periods \citep{van2005multi, cifra2007dynamics, kobayashi2009imaging}, and over several months \citep{jung2005year, laager2008year, cohen1998whole}. In the late 1980s, a Chinese research team led by Yan Zhiqiang detected a more intense photon emission along the meridians of traditional Chinese medicine than that recorded just a few millimeters away \citep{yan1989investigation}. Twenty years later, Yang Joon-Mo et al. conducted an in-depth study of the characteristics of biophoton emission from human hands during several months examining the correlation between emissions from right and left hand, as well as between the palms and the backs of the hands under various conditions (good health, common cold, hemiparesis, during physical exercize \citep{Yang2004biofotonico, jung2003left, laager2008effects}), and in connection with the theory of yin and yang in traditional Chinese medicine \citep{yang2006yin}.

Of direct relevance to the present work, Van Wijk et al. showed that meditation reduces UPE from hands and forehead, and that practitioners of transcendental meditation (TM) exhibit the lowest emission values among the populations studied, with an anatomically specific modulation concentrated at the palmar hollow of the hands \citep{VanWijk2005med, VanWijk2006TM}. Recently, Dyer et al. studied 23 subjects during meditation (loving kindness meditation followed by a guided breath-work exercise):
all of them were equipped with multiple biofield sensors to measure heart rate (HR), heart rate variability (HRV), skin conductance
(SCR), alpha waves with electroencephalography (EEG), infrared radiation (IR), and ultraweak photon emission (UPE) \citep{dyer2026changes}. 

The study of meditative states through physiological signals has grown substantially over the past two decades. 
EEG studies have documented characteristic delta and theta power increases,
increased theta and high-beta functional connectivity, phase-amplitude coupling
between these bands in prefrontal and default-mode-network regions, and
small-world network reorganization during structured nasal breathing practices
\citep{Zaccaro2022}. 
Heart rate variability (HRV) analysis has complemented these findings by revealing that meditation modulates autonomic nervous system dynamics in ways detectable by nonlinear time-series methods. In this context, Diffusion Entropy Analysis (DEA), a method that extracts the scaling exponent $\delta$ from the diffusion entropy generated by a time series, identifying the presence of non-Poissonian, renewal fluctuations known as crucial events, has proven particularly informative. 

Tuladhar et al.\ demonstrated that meditation induces a transition toward
coherent, organized cardiac dynamics, shifting the renewal index $\mu$ of
the inter-beat distribution from values near $\mu \approx 2$ (ideal $1/f$
regime) toward $\mu \approx 3$ (Gaussian basin of attraction), corresponding
to a decrease of the DEA scaling exponent $\delta$ toward the Poisson
reference ($\delta = 0.5$) \citep{Tuladhar2018}. DEA has subsequently been applied to HRV for the detection of autonomic neuropathy, revealing that the loss of complexity is quantifiable through a systematic reduction of $\delta$ \citep{Jelinek2021}. The same analytical framework, developed by our group in collaboration with Grigolini (University of North Texas - UNT) and coworkers, has been successfully extended to biophoton signals from cell cultures and germinating seeds \citep{Benfatto1, DePaolis1, DePaolis2}.

The meditative protocol adopted in the present study is Sama Vritti (Sanskrit: ``equal fluctuation''), commonly known as box breathing or square breathing. It consists of four equal-duration phases , nasal inhalation, breath retention, exhalation, and post-exhalation suspension , each lasting four seconds, yielding a slow, symmetric respiratory cycle at approximately 3.75 cycles per minute. This patterned breathing is known to shift autonomic balance toward parasympathetic dominance and to modulate oscillatory brain activity in low-frequency bands \citep{Zaccaro2022}. We chose Sama Vritti meditation for its simplicity: mental execution only (no mantra to chant); controlled but regular breathing without the risk of hyperventilating; no physical movement (the body
is essentially still throughout).

In this paper we present measurements of UPE from the palm of the right hand of a single subject (E.P.) under four consecutive 15-minute phases: a dark phase (detector shielded), a pre-meditation phase of undirected mind wandering, a Sama Vritti meditation phase, and a post-meditation mind wandering recovery phase. Three independent sessions were conducted on different days. The data are analyzed with a comprehensive battery of complementary methods:
distributional statistics including the Fano factor, skewness, and tail
Expected Shortfall; multiscale Fano factor and Allan deviation;
stripe-filtered DEA; and R\'{e}nyi entropy with a Time Reversal test.

The paper is organized as follows. Section~\ref{sec:apparatus} describes the experimental apparatus and the data collection protocol. Section~\ref{sec:expdata} presents the raw data and basic statistical characterization. Section~\ref{sec:analysis} reports the full complexity analysis. Section~\ref{sec:discussion} discusses the results and draws conclusions. Section~\ref{sec:future} outlines future perspectives.
\vspace{12pt}
\section{Materials and Methods}
\label{sec:apparatus}
\begin{figure}[H]
\centering
\includegraphics[width=0.70 \linewidth]{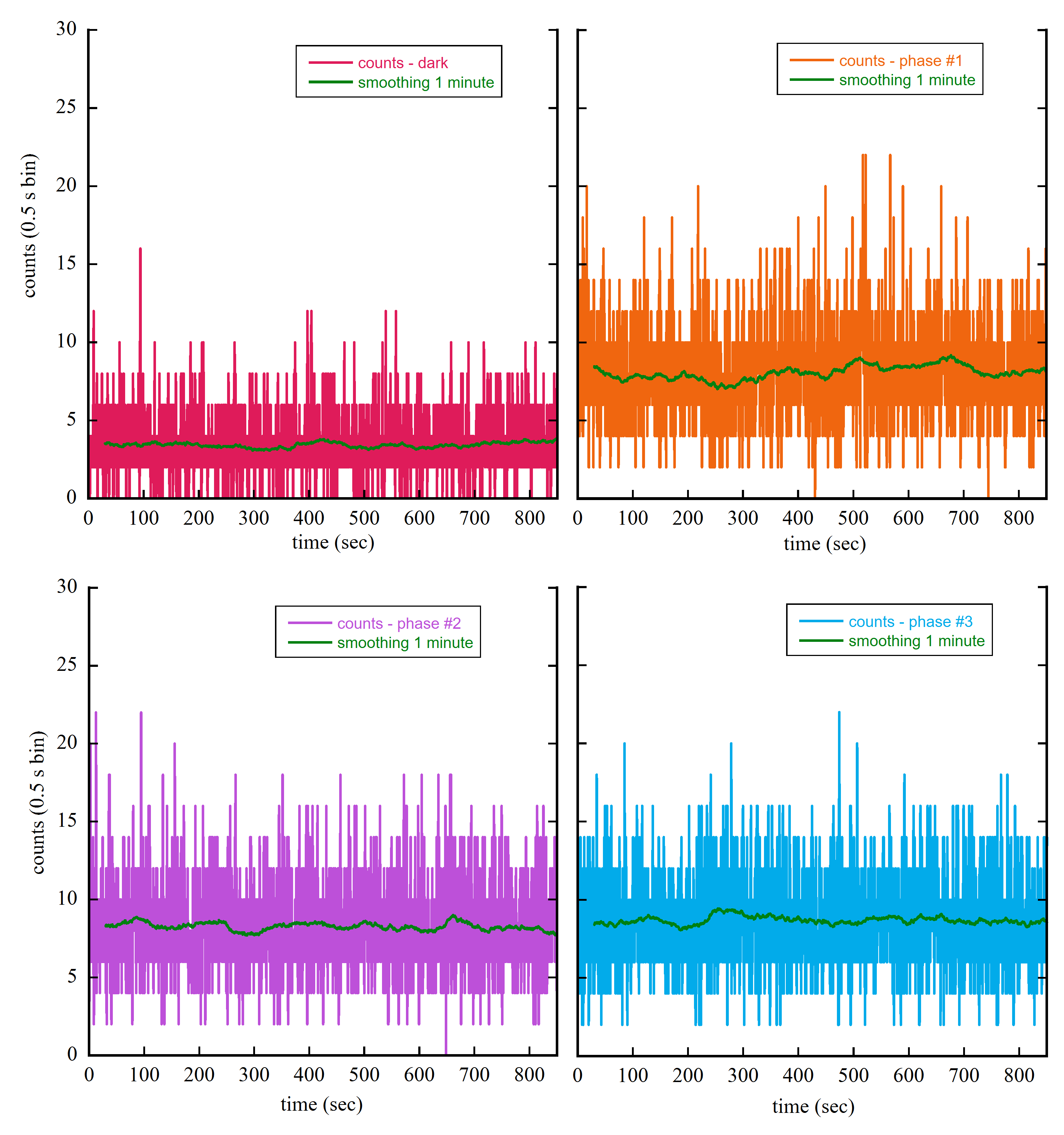}
\caption{\textbf{Representative biophoton count time series (Session~1).}
Each panel shows the photon counts acquired in non-overlapping 0.5~s time bins
over the 900~s duration of one phase. Top left: Dark (phase~\#0); top right:
Pre-meditation (phase~\#1); bottom left: Meditation (phase~\#2); bottom right:
Post-meditation (phase~\#3). The green curve in each panel is a 1-minute running
average. The separation between the Dark baseline and the biological signal
in phases \#1--\#3 is clearly visible.}
\label{fig:rawdata}
\end{figure}

The experimental apparatus employed in this study is a dedicated adaptation of the
photon-counting setup previously developed and validated at LNF-INFN for biophoton
measurements on germinating seeds and cell cultures \citep{Benfatto1, DePaolis1, DePaolis2}.
The core detector is the H12386-210 high-speed photon-counting head (Hamamatsu
Photonics Italia S.r.l., Arese (MI), Italy), operated at +5~V, featuring a circular
active area with a radius of 5~mm and high sensitivity over the 230--700~nm wavelength
range, with a peak response at 400~nm \citep{Benfatto1, DePaolis1}. 
A light-tight enclosure was designed to accommodate a human hand. It is a box with an opening on one side, lined with foam to allow comfortable insertion of the
subject's hand while preserving light isolation. The hand was positioned on a dedicated
support equipped with five aligned rings to guide placement, ensuring that the center
of the palm was consistently aligned with the upward-facing sensitive area of the
photo-counter. An additional light-shielded aperture was provided for routing the
power and signal cables. The entire chamber was covered with black insulating tape
and optically shielded with a double layer of black light-absorbing fabric.
Data acquisition was performed using a custom electronics system developed at
LNF-INFN, which allows the acquisition time window to be set in a controlled manner;
in the present measurements a bin width of 0.5~s was adopted throughout. The acquisition
chain was interfaced with LabVIEW and custom Python-based DAQ software; real-time
visualization was implemented via the TKinter phyton library.
Three independent data acquisition sessions were conducted on different days, each
starting at approximately 11:00~a.m.\ to minimize circadian and ultradian effects. The subject
was E.P.\ (co-author E.~Pace), who wore a thick opaque glove covering her right hand for
several hours prior to each session in order to suppress residual skin luminescence. Before the measurement, the hand was cleaned with a sodium hypochlorite solution (Amuchina\textsuperscript{\textregistered}, an Italian disinfectant brand) to remove surface contaminants that could interfere with or attenuate the emitted signal, and between measurements the experimental setup was always kept closed to prevent any light pollution that
might cause residual luminescence.
Each session lasted
one hour and consisted of four consecutive 15-minute phases. At the start of each
session, the subject's right hand was positioned in the chamber and remained there
without interruption for the entire duration; transitions between phases were
therefore achieved without any mechanical disturbance to the hand position.
The four phases, in fixed order, were as follows: a \textbf{Dark} phase, in which
a light-tight cardboard screen was interposed between the palm and the detector
window to assess the intrinsic detector noise and establish a reference baseline;
a \textbf{Pre-meditation} phase, in which the cardboard was removed , without
displacing the hand , and the subject maintained an undirected resting state
without specific cognitive engagement; a \textbf{Meditation} phase, in which the
subject performed continuous Sama Vritti box breathing (4~s inhalation, 4~s breath
retention, 4~s exhalation, 4~s post-exhalation suspension) for the full 15 minutes;
and a \textbf{Post-meditation} phase, in which the subject returned to an undirected
resting state, maintaining the hand in position. During phases 2 to 4, the subject kept her eyes closed in order to not
introduce any perturbation due to eyes movement or blinking.
A representative example of the raw data from one session is shown in
Figure~\ref{fig:rawdata}.
\FloatBarrier
\section{Experimental Data and Statistical Characterization}
\label{sec:expdata}
Table~\ref{tab:cross_stats} summarises the basic distributional statistics of the
photon count time series for all three sessions and all four acquisition phases.
The mean count per bin $\mu$ clearly separates the Dark phase
($\mu \approx 3.5$--$5.4$ counts/bin) from the three hand phases
($\mu \approx 8$--$12$ counts/bin), confirming a robust biological signal
above the detector baseline in every session.
The overall emission level varies across sessions , a result expected for
a single-subject protocol conducted on different days , but the
internal phase ordering within each session is stable and reproducible.
Notably, the Dark count in Session~3 ($\mu = 5.36$) is anomalously elevated
compared to Sessions~1 and~2, a finding likely attributable to residual
luminescence, and is not reflected in the hand phases, which remain
consistent with the other sessions.
\begin{table}[htbp]
\centering
\begin{tabular}{llcccc}
\toprule
& \textbf{Phase} & $\mu$ & $\sigma$ & $F$ & $\gamma_1$ \\
\midrule
\multirow{4}{*}{Session 1}
  & Dark       &  3.45 & 2.12 & 1.30 & 0.81 \\
  & Pre        &  8.07 & 3.18 & 1.25 & 0.47 \\
  & Meditation &  8.21 & 3.08 & 1.16 & 0.43 \\
  & Post       &  8.63 & 3.12 & 1.13 & 0.27 \\
\midrule
\multirow{4}{*}{Session 2}
  & Dark       &  3.60 & 2.22 & 1.37 & 0.88 \\
  & Pre        & 10.87 & 3.81 & 1.33 & 0.34 \\
  & Meditation & 11.78 & 3.62 & 1.11 & 0.25 \\
  & Post       & 11.32 & 3.71 & 1.21 & 0.44 \\
\midrule
\multirow{4}{*}{Session 3}
  & Dark       &  5.36 & 2.82 & 1.48 & 0.58 \\
  & Pre        &  9.56 & 3.38 & 1.19 & 0.37 \\
  & Meditation & 10.30 & 3.40 & 1.12 & 0.29 \\
  & Post       & 10.45 & 3.57 & 1.22 & 0.40 \\
\bottomrule
\end{tabular}
\caption{Basic distributional statistics across all three sessions and four phases.
$\mu$: mean count per 0.5~s bin; $\sigma$: standard deviation;
$F = \sigma^2/\mu$: Fano factor; $\gamma_1$: empirical skewness.}
\label{tab:cross_stats}
\end{table}
A brief remark is warranted on the absolute level of the Dark counts.
The mean Dark values observed here ($\mu \approx 3.5$--$5.4$ counts per 0.5~s bin,
corresponding to $\approx 7$--$11$ counts/s) are approximately three times higher
than the dark rates recorded in our previous measurements on germinating seeds and
cell cultures, performed with the same phototube at room temperature with 1~s
acquisition per bins \citep{Benfatto1, DePaolis1, DePaolis2}.
We attribute this elevated baseline to two concurrent effects: a modest thermal
increase of the phototube induced by proximity to the subject's hand throughout
the full 60-minute session (including the Dark phase), and residual light entering
the apparatus despite the double light-shielding. The latter interpretation is
supported by the anomalously high Dark value in Session~3 ($\mu = 5.36$), which
suggests a stricter light-isolation condition was not fully maintained in that
session. Importantly, neither effect alters the biological interpretation of the
results, since the hand-phase signals exceed the Dark baseline by a factor of
approximately 2.5--3 in all sessions.
The most robust finding in Table~1 concerns the Fano factor
$F = \sigma^2/\mu$, which quantifies the degree of super-Poissonian fluctuations
relative to a purely random (Poisson) process, for which $F = 1$. In Sessions~2
and~3, $F$ reaches its minimum during the Meditation phase ($F = 1.11$ and $1.12$
respectively), compared to Pre values of $1.33$ and $1.19$. In Session~1, the
reduction initiated at Meditation ($F = 1.16$, down from Pre $= 1.25$) continues
monotonically into Post ($F = 1.13$). Across all three sessions the
Pre$\to$Meditation decrease is therefore consistent and reproducible , despite a
varying overall emission level ($\mu_\mathrm{med} \approx 8.2,\ 11.8,\ 10.3$
counts/bin) , and indicates a genuine compression of count fluctuations associated
with the meditative state. A parallel behaviour is observed in the empirical
skewness $\gamma_1$: in Sessions~2 and~3 its minimum occurs during Meditation
($\gamma_1 = 0.25$ and $0.29$), below the corresponding Pre values ($0.34$ and
$0.37$); in Session~1 the analogous monotonic decrease carries $\gamma_1$ from
Pre $= 0.47$ through Meditation $= 0.43$ to Post $= 0.27$. In Sessions~2 and~3
the Post phase shows a partial rebound of both $F$ and $\gamma_1$ above the
Meditation values; in Session~1 both quantities continue to decrease monotonically
into Post.

This difference likely reflects session-to-session variability in the depth or persistence of the meditative
state, consistent with the expected biological variability of a single-subject protocol. A potential concern
with the fixed phase order is whether the Pre$\to$Meditation reduction of $F$ could reflect a monotone decay
of residual hand luminescence rather than a genuine physiological effect. Two observations argue against
this interpretation. First, the mean count $\mu$ increases from Pre to Meditation in all three sessions
(Table~1), contrary to what a decaying luminescence contribution would produce. Second, Table~1 shows
that in Sessions~2 and 3 the Post phase exhibits a partial recovery of $F$ above the Meditation value
($F$: 1.21 vs 1.11 in Session~2; 1.22 vs 1.12 in Session~3), a non-monotonic behaviour that no
luminescence decay model can account for. In Session~1 the Meditation$\to$Post difference in $F$
(1.16 $\to$ 1.13) is smaller than the expected standard error $F\sqrt{2/N} \approx 0.039$ for
$N \approx 1800$ bins, and is therefore statistically indistinguishable from zero: Session~1 is thus
also consistent with the V-shaped pattern of Sessions~2 and 3. Across all three sessions, the only
statistically robust trend is the Pre$\to$Meditation reduction; the Meditation$\to$Post change is either
a significant rebound (Sessions~2 and 3) or compatible with zero (Session~1), ruling out a monotone
within-session drift in every case. This conclusion is further corroborated by the multiscale Fano
analysis of Section~4.1, where the non-monotonic recovery of $F(\tau)$ across Sessions~2 and 3 is
confirmed at every averaging time scale.

To probe the shape of the count distributions more finely , and in
particular the behaviour of their extreme events , we computed a set of
tail metrics following the procedure introduced in
\citep{DePaolis2, BenfattoArXiv2025}.
For a given quantile $q$, one defines the threshold
$k_q = \min\{k : F_{\rm exp}(k) \geq q\}$,
where $F_{\rm exp}(k) = \sum_{m \leq k} P_m$ is the empirical cumulative
distribution. From this, the right-tail probability mass and Expected
Shortfall are:
\begin{equation}
M_{{\rm dx},q} = \sum_{k \geq k_q} P_k, \qquad
\mathrm{ES}_{{\rm dx},q} = \frac{\sum_{k \geq k_q} k\,P_k}{M_{{\rm dx},q}},
\end{equation}
where $M_{{\rm dx},q}$ measures the probability weight beyond the threshold,
and $\mathrm{ES}_{{\rm dx},q}$ is the conditional mean count given that the
observation exceeds it , a measure of the average intensity of extreme
events rather than just their frequency.
Symmetrically, the left-tail quantities at quantile $q$ are:
\begin{equation}
M_{{\rm sx},q} = \sum_{k \leq k_q} P_k, \qquad
\mathrm{ES}_{{\rm sx},q} = \frac{\sum_{k \leq k_q}(k_q - k)\,P_k}{M_{{\rm sx},q}},
\end{equation}
which capture the probability mass below the threshold and the average gap
between the threshold and the sub-threshold observations.
Table~\ref{tab:cross_tails} reports these quantities at $q = 0.95$
for the three hand phases; 
\begin{table}[H]
\centering
\begin{tabular}{llcccc}
\toprule
& \textbf{Phase}
  & $M_{\rm dx}$ & $M_{\rm sx}$ & $\mathrm{ES}_{\rm dx}$ & $\mathrm{ES}_{\rm sx}$ \\
\midrule
\multirow{3}{*}{Session 1}
  & Pre        & 0.0693 & 0.9769 & 14.968 & 6.119 \\
  & Meditation & 0.0688 & 0.9747 & 14.848 & 5.971 \\
  & Post       & 0.0912 & 0.9714 & 14.639 & 5.558 \\
\midrule
\multirow{3}{*}{Session 2}
  & Pre        & 0.0604 & 0.9709 & 19.182 & 7.380 \\
  & Meditation & 0.0841 & 0.9670 & 18.928 & 6.475 \\
  & Post       & 0.0714 & 0.9703 & 19.185 & 6.966 \\
\midrule
\multirow{3}{*}{Session 3}
  & Pre        & 0.0618 & 0.9718 & 17.018 & 6.676 \\
  & Meditation & 0.0918 & 0.9742 & 16.862 & 5.933 \\
  & Post       & 0.1045 & 0.9588 & 17.232 & 5.899 \\
\bottomrule
\end{tabular}
\caption{Cross-session tail metrics for the three hand phases at quantile $q=0.95$.
$M_{{\rm dx},q}$: right-tail probability mass;
$M_{{\rm sx},q}$: left-tail cumulative mass;
$\mathrm{ES}_{{\rm dx},q}$: right-tail Expected Shortfall (count units);
$\mathrm{ES}_{{\rm sx},q}$: left-tail expected gap below threshold (count units).}
\label{tab:cross_tails}
\end{table}
\noindent The Dark phase is omitted from Table~2 for brevity; rather than reporting its full set
of tail metrics, we summarise here only the most informative quantity for comparison purposes.
For reference, the Dark phase yields $\text{ES}_{\text{dx}}$ values of 8.6, 8.7, and 10.8 counts
in Sessions~1, 2, and~3 respectively, markedly below the range of any biological phase
(14.6--19.2 counts); the elevated value in Session~3 is consistent with its anomalously
high Dark mean already noted in Table~1. This separation confirms the absence of
high-amplitude burst activity in the instrumental background and provides an independent
validation of the Dark/signal distinction.

The tail metrics of Table~\ref{tab:cross_tails} reveal a pattern complementary
to, and fully consistent with, the behaviour of $F$ and $\gamma_1$ in
Table~\ref{tab:cross_stats}.
The right-tail mass $M_{{\rm dx},q}$ does not display a clear or reproducible
ordering across phases: its values are comparable between Pre and Meditation
in most sessions, and the Post phase is sometimes higher (Sessions~1 and~3)
and sometimes not (Session~2). This quantity is sensitive to the exact
location of the discrete quantile threshold, which shifts between sessions
as $\mu$ changes, and should therefore not be interpreted as a stand-alone
indicator.
A more robust picture emerges from the Expected Shortfall.
The right-tail $\mathrm{ES}_{\rm dx}$ decreases systematically from Pre to
Meditation in all three sessions: from $14.97$ to $14.85$, from $19.18$ to
$18.93$, and from $17.02$ to $16.86$ in Sessions~1, 2, and~3 respectively.
This consistent reduction indicates that during meditation the extreme
high-count events are not only slightly less probable but also less intense
on average, pointing to a genuine contraction of the right tail beyond any
simple shift of the distribution mean.
The Post values show session-dependent behaviour , in Session~1 the
contraction continues, while in Sessions~2 and~3 $\mathrm{ES}_{\rm dx}$
partially recovers toward the Pre level. This asymmetry is not surprising:
the physiological transition from structured Sama Vritti breathing back to
undirected mind wandering is unlikely to be instantaneous, and the timescale
of this recovery may vary between sessions , a reflection of the inherent
biological variability of a single-subject protocol.
A symmetric contraction is observed on the left side: $\mathrm{ES}_{\rm sx}$
decreases from Pre to Meditation in all three sessions
($6.12 \to 5.97$, $7.38 \to 6.48$, $6.68 \to 5.93$ in Sessions~1, 2,
and~3), while the Post values again show session-dependent recovery.
The sub-threshold observations are, on average, closer to the threshold
during meditation: the left body of the distribution is also more concentrated.
Taken together, the results of Tables~\ref{tab:cross_stats} and
\ref{tab:cross_tails} converge on a coherent and reproducible picture.
The Fano factor and skewness capture the effect of meditation as a global
reduction of super-Poissonian excess and distributional asymmetry; the
Expected Shortfall resolves it at the level of individual extreme events,
showing that both high and low count bursts become less intense in the
conditional sense. The consistency of this pattern across three independent
sessions , despite day-to-day variability in the absolute emission level
, points to a genuine and reproducible shift of the biophoton emission
process towards a more ordered, less intermittent regime during Sama Vritti
meditation.

The modest absolute magnitude of some of these effects does not diminish their evidential
weight. The four indices reported in Tables~1 and~2 --- $F$, $\gamma_1$, $\text{ES}_{\text{dx}}$,
and $\text{ES}_{\text{sx}}$ --- are not redundant: $F$ quantifies the global variance excess
relative to a Poisson baseline, $\gamma_1$ captures the asymmetry of the count distribution,
while $\text{ES}_{\text{dx}}$ and $\text{ES}_{\text{sx}}$ probe the conditional intensity of
extreme events in the right and left tails respectively, independently of the global moments.
Each index therefore carries information about a distinct aspect of the emission statistics. 
Treating the three sessions as independent replications, a one-sided sign test on the direction
of the Pre$\to$Meditation change gives $p = (1/2)^3 = 0.125$ for each index individually,
where the factor $1/2$ is the probability that a single session shows a decrease by chance
alone --- a result that is not significant in isolation, but reflects the limited number of
sessions rather than the weakness of the effect. To combine the evidence across indices, we
apply Fisher's method under the working assumption of approximate independence: with $k = 4$
sign tests each yielding $p = 0.125$, the combined statistic $X = -2\sum_{i=1}^{k} \ln(p_i)$,
which under Fisher's method follows a $\chi^2(2k) = \chi^2(8)$ distribution, takes the
observed value $X = 16.64$, giving $p = P(\chi^2(8) > 16.64) \approx 0.034$. As is inherent
to any multi-metric analysis on a common dataset, the four indices are not strictly independent,
being derived from the same count series; the combined $p$-value of 0.034 should therefore be
regarded as an indicative lower bound on the true joint significance. The directional
consistency of all four indices across all three sessions remains the primary and most
assumption-free form of evidence.

Beyond the Pre$\to$Meditation direction, the within-session phase ordering provides a further
internal consistency check. In Sessions~2 and~3, both $F$ and $\gamma_1$ display a V-shaped
pattern: their minimum occurs at Meditation and both partially rebound during Post, ruling out
a monotone within-session trend. In Session~1 the Post values continue the downward trajectory,
consistent with a sustained parasympathetic activation beyond the formal end of the meditative
phase, as discussed above. This session-dependent Post behaviour is itself directionally
informative: it implies that the Pre$\to$Meditation reduction is not merely the first step of
a global within-session drift, but a physiologically specific response to the meditative state.

Taken together, the distributional indices of Tables~1 and~2 converge on a coherent and
statistically substantiated picture: Sama Vritti meditation is associated with a genuine,
reproducible shift of the biophoton emission process toward a more ordered, less intermittent
regime, detectable simultaneously in the global moments, the tail structure, and the
within-session phase ordering of the count distribution.

\section{Complexity Analysis}
\label{sec:analysis}
The distributional analysis of Section~\ref{sec:expdata} characterizes the emission
process through its marginal statistics , mean, variance, skewness, and tail metrics
, which collapse all temporal ordering into a single-time probability distribution.
While informative, such descriptors are blind to the dynamical organization of the
signal: the sequential structure of counts, the presence of temporal correlations,
and the way fluctuations evolve across time scales. To probe these aspects, we apply
a battery of three complementary complexity measures. The Fano factor and Allan
deviation (Section~\ref{sec:allan}) characterize the multiscale intermittency of the
photon counting process. Stripe-filtered Diffusion Entropy Analysis
(Section~\ref{sec:dea}) extracts the scaling exponent associated with crucial-event
renewal dynamics in the emission process. R\'{e}nyi entropy
(Section~\ref{sec:renyi}) provides a nonlinear characterization of the count
distribution at multiple orders, capturing structure beyond what the variance alone
retains. Together, these methods provide a multi-dimensional portrait of how the
dynamical organization of UPE changes across physiological states.
\subsection{Fano factor and Allan deviation}
\label{sec:allan}
Let $c_i$ denote the photon counts recorded in the $i$-th bin of duration $\Delta t = 0.5$~s,
and define the aggregated count over a window of $m$ consecutive bins as
$Z_j(\tau) = \sum_{i=(j-1)m+1}^{jm} c_i$, where $\tau = m\,\Delta t$ is the
\emph{averaging time}.
The Fano factor
\begin{equation}
  F(\tau) = \frac{\mathrm{Var}[Z_j(\tau)]}{\langle Z_j(\tau)\rangle}
  \label{eq:fano}
\end{equation}
where $\langle \cdot \rangle$ denotes the sample mean over non-overlapping windows
and $\text{Var}[\cdot]$ the corresponding sample variance, so that
$F(\tau) = \sigma^2_{Z}/\mu_Z$. This quantity 
measures the relative dispersion of event counts at scale $\tau$ and equals unity
for a Poisson process at all scales \citep{LowenTeich2005}. Values above unity signal
super-Poissonian clustering; the growth of $F(\tau)$ with $\tau$ reflects the
presence of long-range temporal correlations in the emission dynamics, and a
power-law scaling $F(\tau) \sim \tau^{\alpha_F}$ with $\alpha_F > 0$ is the
signature of a process with structured intermittency across scales \citep{LowenTeich2005}.
The complementary Allan deviation
\begin{equation}
  \sigma_A^2(\tau) = \tfrac{1}{2}
  \bigl\langle [\bar{x}_{i+1}(\tau) - \bar{x}_i(\tau)]^2 \bigr\rangle_i,
  \label{eq:allan}
\end{equation}
where $\bar{x}_i(\tau)$ is the local mean of $x(t)$ over the $i$-th interval of
duration $\tau$, probes the temporal stability of the signal and is related to
the power spectral density through a band-pass filter centered at $f \sim 1/\tau$
\citep{Allan1966}. For white (shot) noise $\sigma_A \propto \tau^{-1/2}$; deviations
from this reference slope indicate correlated fluctuations. In log-log
representation, fitting $\sigma_A(\tau) \propto \tau^{\alpha_A}$ yields the
exponent $\alpha_A$, which for fractional Gaussian noise with Hurst exponent $H$
satisfies $\alpha_A = H - 1$.

The Fano factor $F(\tau)$ is systematically lowest during Meditation at every
scale in all three sessions, without exception. The separation from the Pre phase
grows with $\tau$: at $\tau = 0.5$~s the differences are modest
(Pre: 1.22--1.36; Med: 1.12--1.19), while by $\tau = 8$~s the Pre phase in
Sessions~2 and~3 shows pronounced clustering ($F = 3.49$ and $1.88$ respectively)
while Meditation remains near unity ($F = 1.41$ and $1.18$). The values of $F$
at $\tau = 0.5$~s reported here are obtained from the non-overlapping window
aggregation of the multiscale procedure and may differ slightly from those in
Table~1, which are computed directly as $F = \sigma^2/\mu$ from the full series;
the small discrepancy is a finite-sample numerical artefact with no physical
significance.

\begin{figure}[htbp]
\centering
\includegraphics[width=0.72\linewidth]{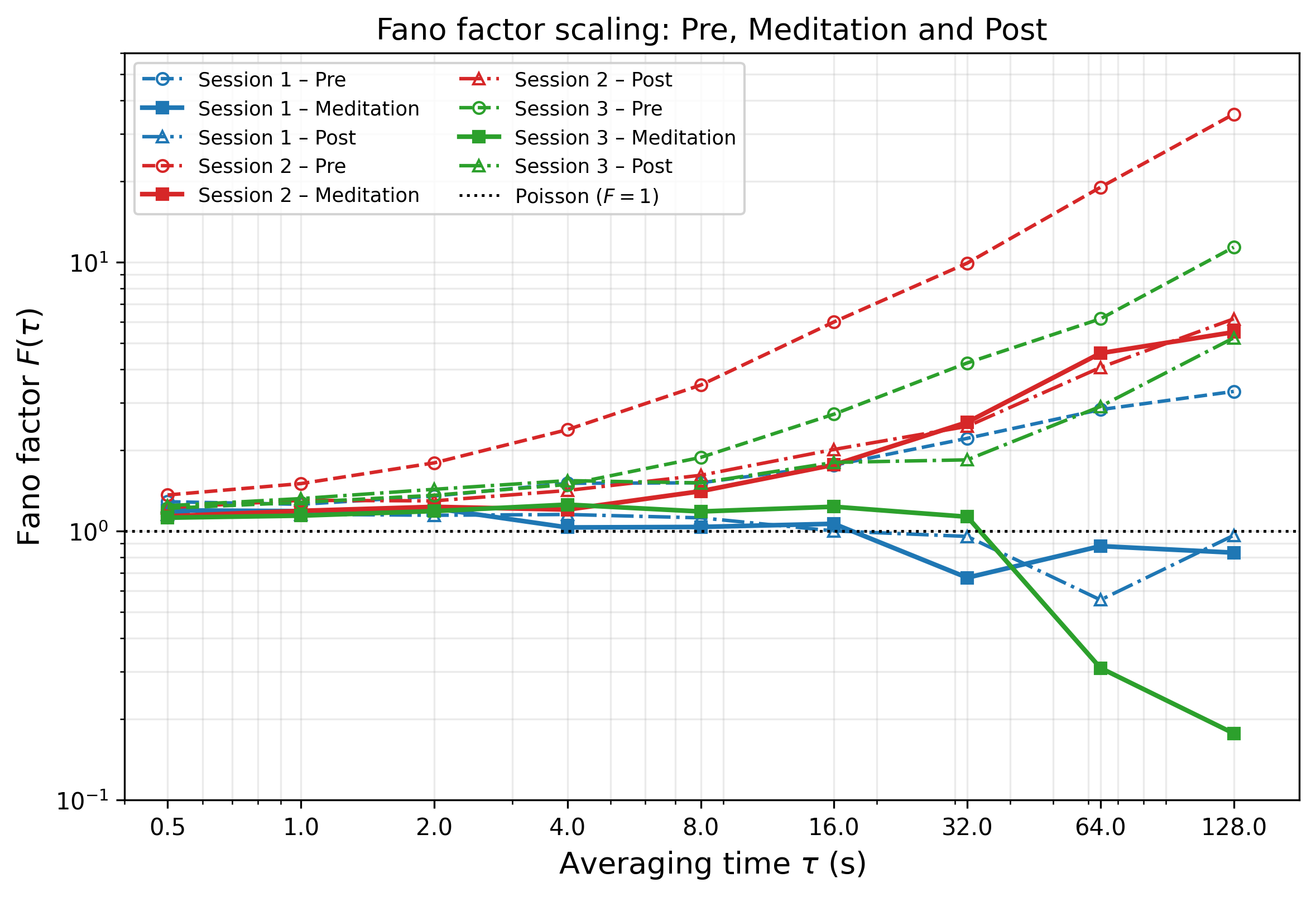}
\caption{Fano factor $F(\tau)$ as a function of averaging time $\tau$
for the Pre-meditation (dashed), Meditation (solid), and Post-meditation
(dash-dot) phases in the three experimental sessions (blue: Session~1;
red: Session~2; green: Session~3). The horizontal dotted line marks the
Poisson reference $F = 1$. In all sessions the Pre phase shows a positive
scaling $F(\tau) \propto \tau^{\alpha_F}$ with $\alpha_F > 0$, reflecting
long-range clustering of the emission process. During Meditation the
scaling becomes flat or negative, indicating a genuine suppression of
multiscale intermittency; the Pre$\to$Meditation difference in $\alpha_F$
is statistically significant at $4.3$--$8.0\sigma$ across the three
sessions. The Post phase shows a partial recovery toward Pre levels in
Sessions~2 and~3, while in Session~1 the suppression persists throughout
the Post phase, reflecting the session-dependent timescale of recovery
from the meditative state.}
\label{fig:fano_pre_med}
\end{figure}

This behaviour is illustrated in Figure~\ref{fig:fano_pre_med}, which shows
$F(\tau)$ in log-log scale for the Pre, Meditation, and Post phases of all three
sessions. The Pre curves (dashed lines) rise consistently across scales, with a
slope that steepens from Session~1 to Session~3; the Meditation curves (solid
lines) are comparatively flat and converge toward, or fall below, the Poisson
reference $F = 1$ at intermediate and large scales. The Post curves (dash-dot
lines) display a session-dependent pattern: in Sessions~2 and~3, $F(\tau)$
rebounds toward Pre levels during Post but does not recover fully within the
15-minute window, indicating a finite relaxation timescale for the physiological
transition back from structured breathing to undirected mind wandering; in
Session~1 the Post phase continues the suppression initiated during Meditation,
consistent with the persistence of parasympathetic activation beyond the formal
end of the meditative phase.
A power-law fit $F(\tau) \propto \tau^{\alpha_F}$ to the full range of scales
confirms this picture: $\alpha_F$ is positive and well-determined for the Pre
phase in all three sessions ($R^2 \geq 0.90$, values $+0.18$, $+0.60$, $+0.40$),
while the Meditation phase shows a markedly reduced or negative exponent
($-0.08$, $+0.29$, $-0.28$). The Pre$\to$Meditation difference is statistically
significant at $8.0\sigma$, $4.3\sigma$, and $5.9\sigma$ in Sessions~1, 2,
and~3 respectively. However, the goodness of the fit varies substantially across
phases, with several $R^2 < 0.5$, so these exponents serve as a qualitative
confirmation of the visual trend in Figure~\ref{fig:fano_pre_med} rather than
as primary quantitative indices. This session-dependent recovery behaviour of
the Post phase is incompatible with a monotone within-session luminescence decay,
which would produce a strictly decreasing trend across all four phases,
corroborating the argument presented in Section~3.

\begin{figure}[H]
\centering
\includegraphics[width=0.60\linewidth]{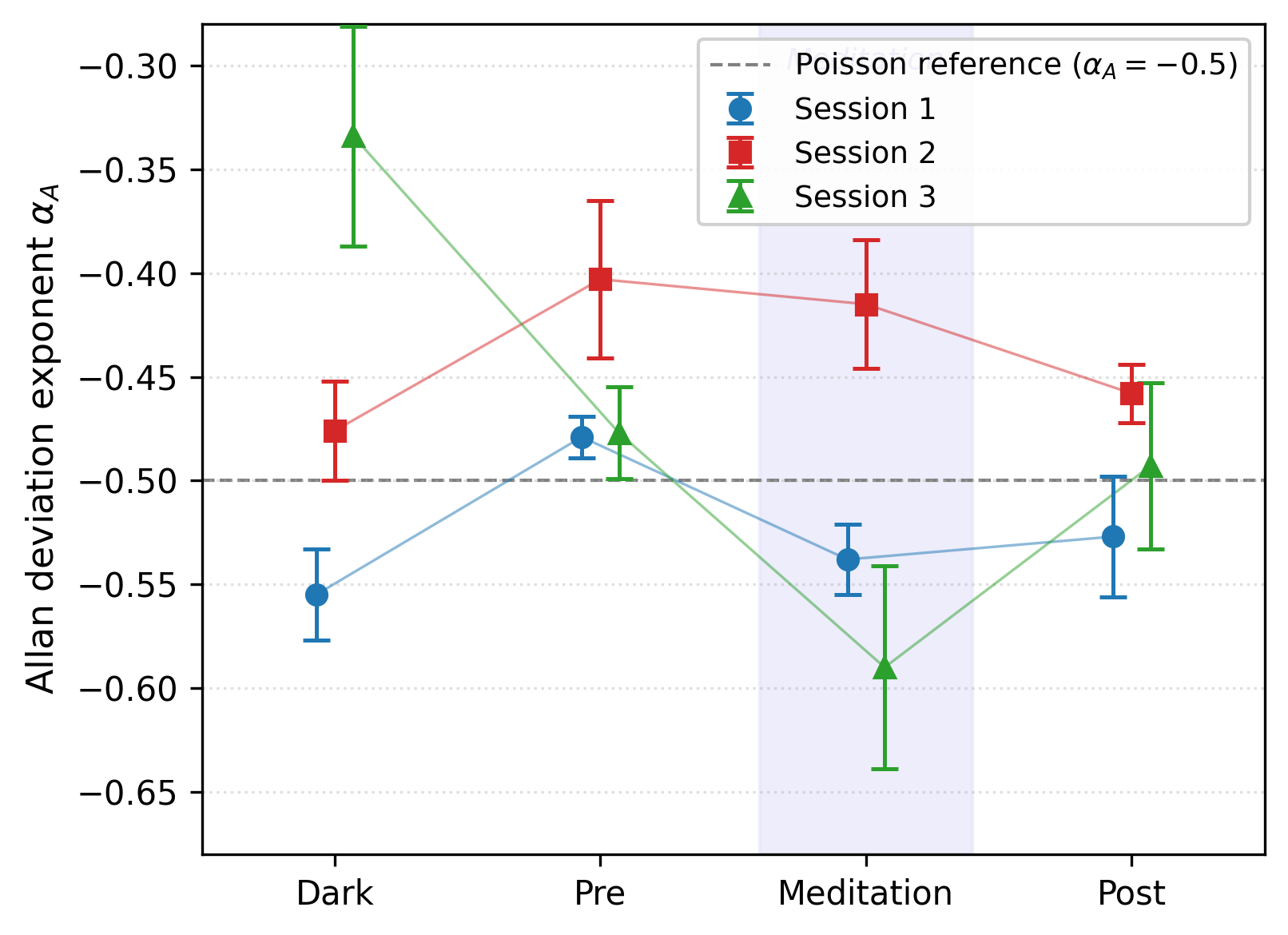}
\caption{Allan deviation exponent $\alpha_A$ vs.\ experimental phase 
(blue circles: Session~1; red squares: Session~2; green triangles: 
Session~3). Error bars: $1\sigma$ from the log-log fit over 
$\tau = 0.5$--128~s. Dashed line: Poisson reference $\alpha_A = -0.5$; 
shaded band: Meditation phase. The Pre$\to$Meditation differences are 
$+0.059$ ($3.0\sigma$), $+0.012$ ($0.2\sigma$), and $+0.113$ ($2.1\sigma$) 
in Sessions~1, 2, and~3}
\label{fig:allan_alpha}
\end{figure}

A more robust characterization is provided by the Allan deviation exponent
$\alpha_A$, obtained by fitting $\sigma_A(\tau) \propto \tau^{\alpha_A}$.
Figure~\ref{fig:allan_alpha} shows the results; all fits have $R^2 \geq 0.94$
except Dark Session~3 ($R^2 = 0.85$), consistent with its anomalous baseline.
The Poisson reference is $\alpha_A = -0.5$. 
For a Poisson process, in which successive counts are statistically independent,
the variance of the mean over an interval of duration $\tau$ scales as $1/\tau$,
so that the Allan deviation scales as $\sigma_A(\tau) \propto \tau^{-1/2}$, giving
$\alpha_A = -0.5$. Values less negative than $-0.5$ indicate the presence of
positive temporal correlations persisting across consecutive intervals, which slow
the decay of $\sigma_A(\tau)$ relative to the Poisson reference; values below
$-0.5$, by contrast, signal anti-correlated fluctuations in which consecutive
local means tend to compensate each other, a signature of active regularisation
of the emission process.
The Pre phase is consistently the least
negative among the biological phases in all three sessions ($\alpha_A = -0.479$,
$-0.403$, $-0.477$), indicating stronger temporal correlations. The Meditation phase yields the most negative exponent in Sessions~1 and~3
($-0.538$ and $-0.590$ respectively); in Session~2 the Post phase is marginally
more negative than Meditation ($-0.458$ vs $-0.415$, a difference of $1.3\,\sigma$),
so the two phases are indistinguishable within the combined uncertainties;
Session~3 is the only case falling measurably below $-0.5$, a signature of
anti-correlated, actively regularized fluctuations. The Pre$\to$Meditation
differences reach $3.0\sigma$ in Session~1 and $2.1\sigma$ in Session~3; in
Session~2 the difference is negligible ($0.2\sigma$), indicating that in that
session the regularization occurs primarily at short scales , captured by the
Fano factor , rather than in consecutive-interval correlations probed by the
Allan deviation. The two indices are therefore genuinely complementary.

\subsection{Diffusion Entropy Analysis with stripe filter}
\label{sec:dea}
Diffusion Entropy Analysis (DEA) is a method for detecting long-range scaling
in time series without the statistical bias introduced by heavy tails or
non-stationarity that affects variance-based estimators such as in Detrended Fluctuation Analysis.
\citep{Scafetta1, Allegrini1, Allegrini2}.
Starting from the raw time series $\{c_i\}$ of photon counts per bin,
one constructs a diffusion trajectory
\begin{equation}
  x(t) = \sum_{i=1}^{t} c_i,
  \label{eq:dea_diffusion}
\end{equation}
and generates an ensemble of realizations by sliding a window of length $l$:
\begin{equation}
  x(l, t) = \int_{t}^{t+l} c(t')\,dt'.
  \label{eq:dea_window}
\end{equation}
The Shannon entropy of the probability density $p(x, l)$ of these realizations,
\begin{equation}
  S(l) = -\int_{-\infty}^{+\infty} p(x, l)\,\ln p(x, l)\,dx,
  \label{eq:dea_entropy}
\end{equation}
grows in the scaling regime as
\begin{equation}
  S(l) = A + \delta\,\ln(l),
  \label{eq:dea_scaling}
\end{equation}
where the slope $\delta$ characterizes the nature of the temporal complexity
in the series \citep{Scafetta1, Grigolini1}.
For a Poisson process which represents the reference of uncorrelated memoryless emission, the scaling factor is $\delta = 0.5$.
When complexity arises from \emph{crucial events} , that is, renewal events drawn from
a waiting-time distribution $\psi(\tau) \propto \tau^{-\mu}$ with $1 < \mu < 3$, which
reset the system memory at each occurrence and generate long-range temporal organization
\citep{Grigolini1, Culbreth} , the relationship between $\delta$ and $\mu$
takes three distinct forms \citep{Grigolini1, Scafetta1}:
\begin{equation}
  \delta =
  \begin{cases}
    \mu - 1 & \text{if}\quad 1 < \mu < 2, \\[4pt]
    \dfrac{1}{\mu - 1} & \text{if}\quad 2 < \mu < 3, \\[6pt]
    0.5 & \text{if}\quad \mu > 3.
  \end{cases}
  \label{eq:delta_mu}
\end{equation}
Values $\delta > 0.5$ therefore indicate a process with $\mu < 3$, i.e., renewal
fluctuations carrying genuine long-range temporal organization beyond ordinary statistical
physics.
The intermediate regime $2 < \mu < 3$ is of particular biological relevance, corresponding
to a non-stationary, non-ergodic process in which the statistical weight of large
inter-event intervals is sufficiently heavy to generate anomalous diffusion, yet the
variance of the waiting-time distribution remains finite \citep{Grigolini1, Culbreth}. 
This regime has been associated with healthy biological dynamics in other 
physiological signals, such as heart rate variability, where disease 
progression drives $\mu$ toward 3, approaching ordinary statistical 
behavior~\citep{Jelinek2021}.

\begin{figure}[htbp]
\centering
\includegraphics[width=0.60\linewidth]{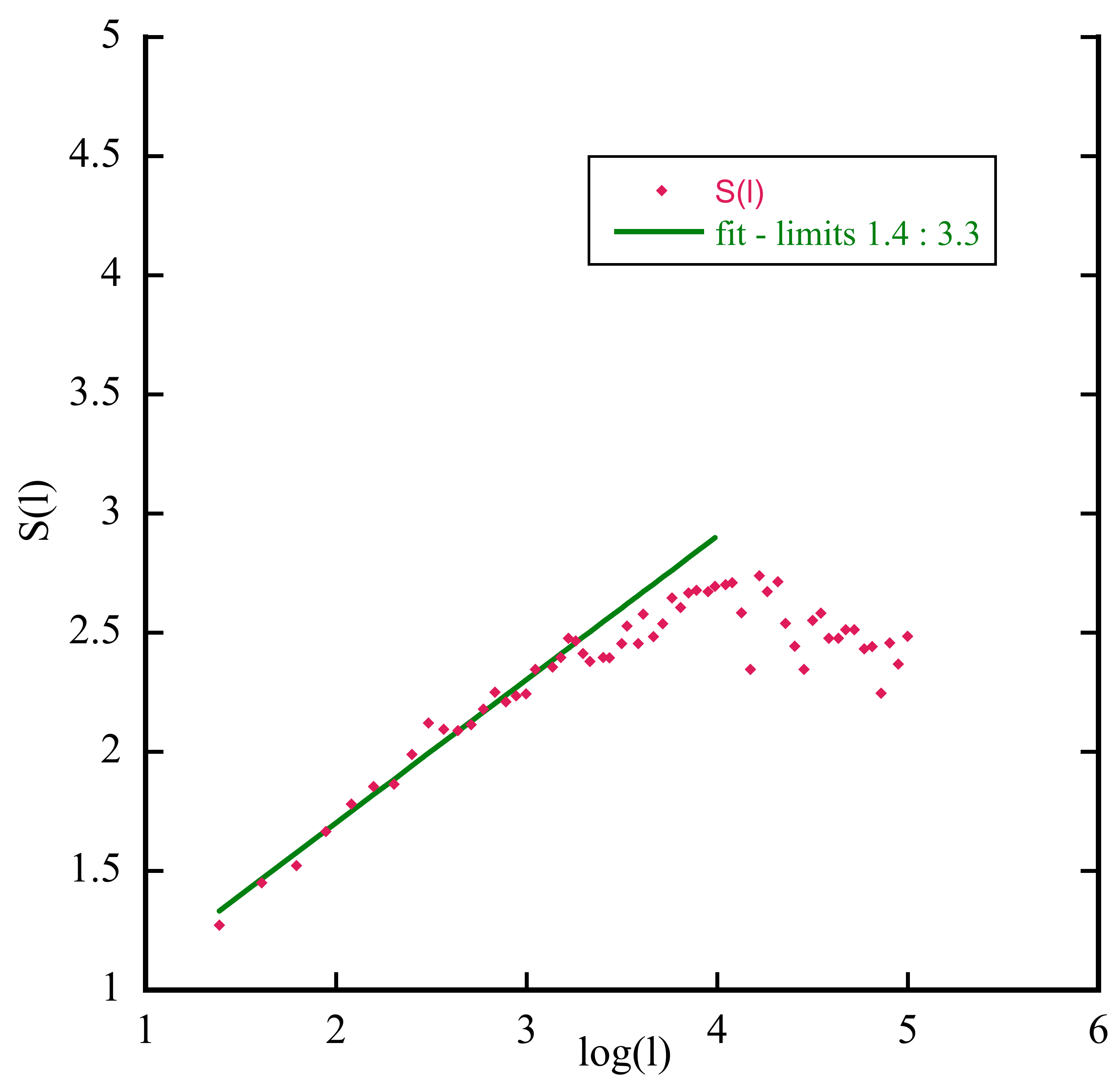}
\caption{Stripe-filter DEA entropy $S(l)$ as a function of $\ln(l)$ for the
Pre-meditation phase of Session~2 ($\delta = 0.602 \pm 0.017$).
Pink diamonds: measured entropy; green line: linear fit
$S(l) = A + \delta\,\ln(l)$ performed over the range $\ln(l) \in [1.4, 3.3]$.
At window sizes beyond $\ln(l) \approx 3.3$ the number of independent
realizations becomes too small relative to the series length ($\sim$1800 bins),
producing the visible fluctuations and eventual saturation of $S(l)$.
Notwithstanding these finite-series effects, the linear scaling region is
clearly identifiable and the fit slope is stable within the selected range,
yielding $\delta > 0.5$ and confirming the presence of
renewal-type correlations in the emission.}
\label{fig:dea_example}
\end{figure}

When DEA is applied directly to the raw biophoton count series, the Poissonian
background dominates the diffusion and the estimated $\delta$ is pulled toward $0.5$,
concealing any underlying critical structure.
To circumvent this, we adopt the stripe-filter modification of DEA \citep{Allegrini1},
originally introduced to separate crucial-event dynamics from non-crucial FBM-type
complexity.
Rather than using the raw counts as diffusion increments, the ordinate axis of the
signal is partitioned into equal-width stripes of size $s$; an event is registered each
time the signal crosses from one stripe into a neighboring one, producing a surrogate
binary series $z(t) = 1$ at crossing times and $z(t) = 0$ otherwise.
The diffusion trajectory of Eq.~(\ref{eq:dea_diffusion}) is then rebuilt using $z(t)$
in place of $c_i$, and the entropy scaling of Eq.~(\ref{eq:dea_scaling}) is evaluated
on this surrogate.
This operation suppresses the Poissonian noise contribution , which produces rapid,
uncorrelated crossings concentrated at the shot-noise scale , while preserving the
slow, correlated crossing events that carry the genuine renewal dynamics of the
biological signal.
In our implementation, the stripe width is set to
$s \simeq 3\sigma_c$, where $\sigma_c$ is the standard deviation of the count series
in the phase under analysis; this empirical choice was introduced to suppress the
Poissonian noise contribution in biophoton measurements and validated on the same
instrumental setup in \citep{BenfattoArXiv2025, DePaolis2}.
Table~\ref{tab:dea_stripes} reports the stripe-filter DEA exponent $\delta$ and its
uncertainty $\sigma_\delta$ for all three sessions and four phases.
An illustrative example of the entropy scaling $S(l)$ vs.\ $\ln(l)$ is shown in
Figure~\ref{fig:dea_example}.
\begin{table}[htbp]
\centering
\begin{tabular}{llcc}
\toprule
& \textbf{Phase} & $\delta$ & $\sigma_\delta$ \\
\midrule
\multirow{4}{*}{Session 1}
  & Dark       & 0.446 & 0.015 \\
  & Pre        & 0.580 & 0.038 \\
  & Meditation & 0.531 & 0.038 \\
  & Post       & 0.521 & 0.017 \\
\midrule
\multirow{4}{*}{Session 2}
  & Dark       & 0.490 & 0.025 \\
  & Pre        & 0.602 & 0.017 \\
  & Meditation & 0.470 & 0.048 \\
  & Post       & 0.540 & 0.016 \\
\midrule
\multirow{4}{*}{Session 3}
  & Dark       & 0.494 & 0.042 \\
  & Pre        & 0.551 & 0.039 \\
  & Meditation & 0.581 & 0.014 \\
  & Post       & 0.659 & 0.030 \\
\bottomrule
\end{tabular}
\caption{Stripe-filter DEA scaling exponent $\delta$ and its uncertainty $\sigma_\delta$
for the three experimental sessions and four phases. The Poisson reference is
$\delta = 0.5$; values above this threshold signal the presence of renewal-type
critical fluctuations. The stripe width was set to $s \simeq 3\sigma_c$ for each
phase independently \citep{BenfattoArXiv2025}. All biological phases satisfy this condition
in all three sessions, consistent with previous findings on cell cultures and
germinating seeds.}
\label{tab:dea_stripes}
\end{table}
A robust feature of Table~\ref{tab:dea_stripes} is that the Dark phase yields the
lowest $\delta$ in all three sessions ($\delta = 0.446$, $0.490$, $0.494$ in
Sessions~1, 2, and~3 respectively).
These numbers are consistent within statistical fluctuations with the Poisson reference, consistent with the
predominantly memoryless character of instrumental background noise; 
Using Eq.~(\ref{eq:delta_mu}), the Dark exponents correspond to $\mu \geq 3$,
confirming the absence of crucial-event dynamics in the detector background.
The biological phases , Pre, Meditation, and Post , yield $\delta > 0.5$ in the
majority of cases, with values $\delta \approx 0.52$--$0.66$ corresponding to
$\mu \approx 2.5$--$2.9$ via the $2 < \mu < 3$ branch of Eq.~(\ref{eq:delta_mu}).
This places the hand emission dynamics in the non-ergodic, non-stationary renewal regime,
consistent with the findings from cell cultures and germinating seeds reported by our
group \citep{Benfatto1, DePaolis1, DePaolis2}.

The Pre~$\to$~Meditation transition does not follow a uniform directional pattern.
Session~1 shows a modest decrease ($\Delta\delta = -0.049$), within the quadrature
sum of the individual uncertainties (0.054). Session~2 shows a more significant
decrease ($\Delta\delta = -0.132$, approximately $2.6\sigma$), indicating a genuine
reduction of the renewal correlation strength during the meditative state. Session~3,
by contrast, shows a slight increase ($\Delta\delta = +0.030$, well within the errors).
The Pre$\to$Meditation decrease of $\delta$ constitutes the central result of the DEA
analysis, consistent with the global picture of reduced dynamical intermittency
established by the Fano factor and Allan deviation. To combine the evidence across
$k = 3$ sessions, we apply Stouffer's method to the individual z-scores:
$z_1 = -0.049/0.054 \approx -0.91$, $z_2 = -0.132/0.051 \approx -2.60$,
$z_3 = +0.030/0.041 \approx +0.73$, giving
\begin{equation}
z_{\text{comb}} = \frac{z_1 + z_2 + z_3}{\sqrt{k}} = \frac{-2.78}{\sqrt{3}} \approx -1.60.
\end{equation}
Under Stouffer's method $z_{\text{comb}}$ follows a standard normal distribution;
the observed value $z_{\text{comb}} \approx -1.60$ therefore gives
$p = P(Z < -1.60) \approx 0.055$ (one-tailed), where $Z \sim \mathcal{N}(0,1)$.
This result approaches conventional significance: two sessions show a decrease (one
significant at $2.6\sigma$, one marginal), and the increase observed in Session~3 is
indistinguishable from zero within the measurement uncertainty, so that the overall
directional evidence points consistently toward a genuine reduction of renewal
correlation strength during the meditative state.
The session-to-session variability
of this result has two non-separable origins. The first is methodological: with
15~min of data at 0.5~s binning, the DEA scaling range spans less than one decade
in window length, making $\delta$ intrinsically sensitive to the fit boundaries and
to the number of stripe-crossing events realised per phase. The second is genuine
biological variability: the depth of the meditative state and the physiological
arousal level of the subject on a given day are not controlled parameters in a
single-subject protocol.
\subsection{R\'{e}nyi entropy}
\label{sec:renyi}
The R\'{e}nyi entropy of order $\alpha$ of a discrete probability distribution
$\{p_k\}$ over $M$ outcomes is defined as
\begin{equation}
  H_\alpha(\{p_k\}) = \frac{1}{1-\alpha}\,\ln\!\left(\sum_{k=1}^{M} p_k^\alpha\right),
  \qquad \alpha \neq 1,
  \label{eq:renyi}
\end{equation}
with the limiting case $\lim_{\alpha\to 1}H_\alpha = -\sum_k p_k\ln p_k$
recovering the Shannon entropy.
The parameter $\alpha$ acts as a continuous sensitivity dial: for $\alpha < 0$,
the sum $\sum_k p_k^\alpha$ diverges for $p_k \to 0$, so $H_\alpha$ is dominated
by the outcomes with the \emph{smallest} probabilities and is therefore sensitive
to rare, high-amplitude events in the tail of the distribution; for $\alpha > 0$,
larger probabilities are weighted more heavily and $H_\alpha$ probes the
\emph{bulk}. The Shannon limit $\alpha = 1$ treats all outcomes with equal
logarithmic weight and is thus insensitive to the tail--bulk distinction.
Computing $H_\alpha$ over a range $\alpha \in [-8, 6]$ on a grid of 501 equally
spaced values therefore yields a curve $H_\alpha(\alpha)$ that encodes the full
distributional shape, from tail-dominated to bulk-dominated regimes.
The absolute scale of $H_\alpha$ depends on the number of outcomes $M$, which
differs between the two methods described below. To obtain curves that are
directly comparable across sessions and phases, each method uses its own
normalisation by the corresponding maximum-entropy (Hartley) reference
$H_\alpha^{\rm max} = \ln M$, which is the entropy of the uniform distribution
over $M$ outcomes and is the maximum possible value of $H_\alpha$ for any
$\alpha > 0$. The normalised curve is
\begin{equation}
  \tilde{H}_\alpha = \frac{H_\alpha}{H_\alpha^{\rm max}},
  \label{eq:renyi_norm}
\end{equation}
so that $\tilde{H}_0 = 1$ by construction for every distribution, anchoring
all curves at a common reference at $\alpha = 0$.
A curve that lies \emph{above} this anchor at $\alpha < 0$ indicates heavier
tails than the uniform reference; one that falls steeply for $\alpha > 0$
signals a concentrated, peaked bulk.
\paragraph{Direct method.}
The first approach estimates the distribution $\{p_k\}$ of photon counts per
bin directly from the data.
The raw count series is first linearly normalised to the unit interval $[0,1]$
and a probability density is estimated by Gaussian kernel density estimation
(KDE) with bandwidth chosen by Scott's rule.
To avoid the well-known boundary bias of KDE near $x = 0$ and $x = 1$,
the density is computed using the reflection method: artificial mirror copies
of the data are concatenated at $x < 0$ and $x > 1$ before fitting, so that
the fitted density integrates correctly near the boundaries.
The density is then evaluated on a uniform grid of $N_{\rm KDE} = 1000$ points
in $[0,1]$, and the discrete probability vector $\{p_k\}_{k=1}^{1000}$ is
obtained by normalising the resulting bin areas.
The normalisation factor for this method is therefore
$H_\alpha^{\rm max} = \ln N_{\rm KDE} = \ln 1000 \approx 6.91$,
corresponding to the maximum entropy over $N_{\rm KDE}$ equally probable
bins.
This approach is insensitive to temporal ordering: it probes only the
\emph{marginal} amplitude distribution $p(I)$ and is blind to any
sequential structure in the emission process.
The area-tilt scalar $T_{\rm dir}$ (defined below in Eq.~\ref{eq:tilt})
is extracted from $\tilde{H}^{\rm dir}_\alpha$.
\paragraph{Sequence method.}
The second approach targets the \emph{joint} distribution of successive
emission states and is therefore sensitive to temporal correlations.
The time series $\{c_i\}$ is first converted into a symbolic sequence
$\{s_i\}$ by quantile discretisation: the $N \approx 1800$ count values are
ranked and divided into $K = 4$ equal-frequency bins (quartiles), assigning
to each bin a symbol $s_i \in \{1,2,3,4\}$.
Consecutive triples $(s_i, s_{i+1}, s_{i+2})$ form \emph{3-grams}
(length-$L=3$ patterns); there are $K^L = 4^3 = 64$ distinct patterns.
The empirical frequency of each pattern over the $N - L + 1 \approx 1797$
overlapping triplets in the series defines the probability distribution
$\{p_{\mathbf{s}}\}$ over the 64-element pattern space.
The normalisation factor for the sequence method is the maximum entropy of
this pattern space:
$H_\alpha^{\rm max} = \ln(K^L) = L\ln K = 3\ln 4 \approx 4.16$,
corresponding to the uniform distribution over 64 equally probable patterns.
With $\sim$1800 bins and 64 patterns, each pattern is observed on average
$\sim$28 times, ensuring statistically reliable empirical probabilities.
A coarser symbolisation ($K = 16$, $K^L = 4096$, $\sim$0.3 observations
per pattern) was tested and found to produce results dominated by sampling
noise; those results are not reported.
The key property distinguishing this method from the direct approach is that
two processes with \emph{identical} marginal distributions but \emph{different}
temporal correlations , for instance an i.i.d.\ process and a Markov chain
with the same stationary distribution , will in general produce distinct
pattern distributions and thus distinct $\tilde{H}^{\rm seq}_\alpha$ curves.
The sequence method therefore adds sensitivity to the short-range temporal
organisation of the signal that is inaccessible to the direct method.
The area-tilt scalar $T_{\rm seq}$ is extracted from $\tilde{H}^{\rm seq}_\alpha$.
\paragraph{Area-tilt index.}
To reduce each normalised R\'{e}nyi curve to a single scalar that quantifies
the tail--bulk asymmetry, we define the \emph{area-tilt} index:
\begin{equation}
  T = \frac{\langle\tilde{H}\rangle_{[-6,\,0]} - \langle\tilde{H}\rangle_{[0,\,6]}}
      {\langle\tilde{H}\rangle_{[-6,\,0]} + \langle\tilde{H}\rangle_{[0,\,6]}},
  \label{eq:tilt}
\end{equation}
where $\langle\tilde{H}\rangle_{[a,b]}$ denotes the trapezoidal-rule mean of
$\tilde{H}_\alpha$ over the 501-point grid restricted to $[a,b]$.
Since $\tilde{H}_0 = 1$ for all distributions, both averages are anchored
at the same value at $\alpha = 0$; a distribution with heavier tails will
produce a higher $\tilde{H}_\alpha$ for $\alpha < 0$ and hence a larger
numerator, yielding $T > 0$.
A purely Poissonian process at the count rates observed here gives
$T_{\rm dir} \approx 0.13$--$0.19$ (Dark phases); lower values in the
active phases indicate a distribution with relatively lighter tails.
\paragraph{Time Reversal test.}
As a third index that specifically targets the \emph{arrow of time} in the
sequential dynamics, we compute the Jensen--Shannon divergence between the
3-gram distribution of the original (forward) series and that of its
time-reversed counterpart.
Given two distributions $P = \{p_{\mathbf{s}}\}$ and $Q = \{q_{\mathbf{s}}\}$
over the same pattern space, the Kullback--Leibler divergence is
$D_{\rm KL}(P\|Q) = \sum_{\mathbf{s}} p_{\mathbf{s}} \ln(p_{\mathbf{s}}/q_{\mathbf{s}})$,
and the Jensen--Shannon divergence is the symmetrised version:
\begin{equation}
  J_{\rm TR} = D_{\rm JS}(P_{\rm fwd} \| P_{\rm rev})
  = \tfrac{1}{2}D_{\rm KL}(P_{\rm fwd}\|M) + \tfrac{1}{2}D_{\rm KL}(P_{\rm rev}\|M),
  \label{eq:jtr}
\end{equation}
where $M = \tfrac{1}{2}(P_{\rm fwd}+P_{\rm rev})$ is the pointwise mixture
distribution and $P_{\rm rev}$ is obtained by reversing the entire time series
before extracting 3-grams.
By construction $0 \leq J_{\rm TR} \leq \ln 2$; for a stationary
time-reversible process, $P_{\rm fwd} = P_{\rm rev}$ and $J_{\rm TR} = 0$
exactly. Non-zero values signal \emph{broken time-reversal symmetry}, a
signature of non-equilibrium dynamics sustained by continuous energy
dissipation \citep{Grigolini1}.
To ensure adequate pattern sampling for this test independently of each
phase, the symbolisation uses a number of quantile bins $K_{\rm tr}$
computed so that the expected number of 3-gram observations per pattern is
at least 20; $K_{\rm tr} \leq K = 4$ and equals 4 in all phases analysed here.

\begin{figure}[t]
\centering
\includegraphics[width=0.80\linewidth]{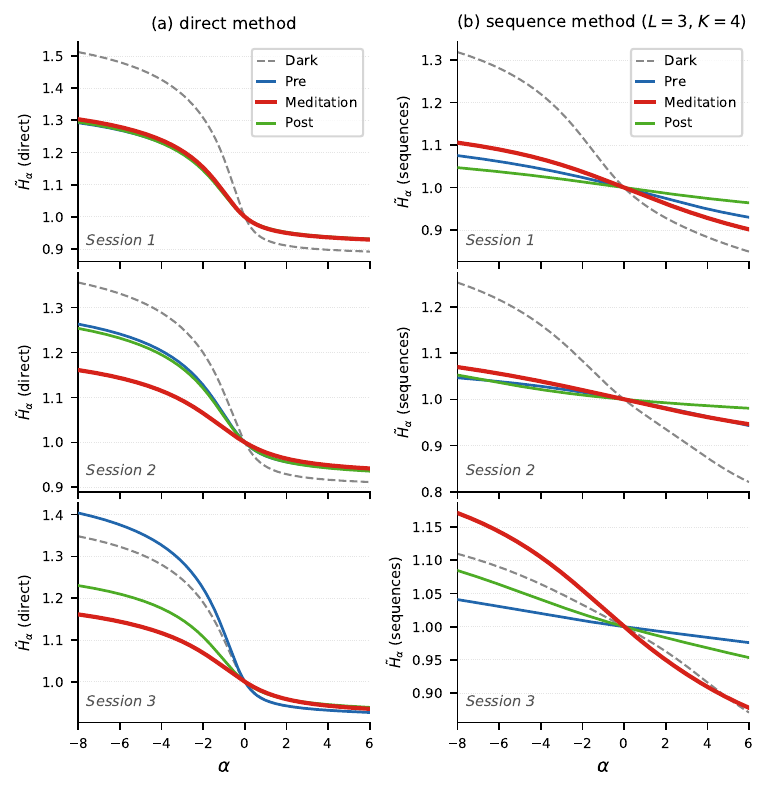}
\caption{Normalised R\'{e}nyi entropy $\tilde{H}_\alpha(\alpha)$ for the
three experimental sessions and four phases. Left column: direct KDE method.
Right column: sequence method ($L=3$, $K=4$ quantile symbolisation).
Rows from top to bottom: Sessions~1, 2, and~3. Vertical dotted lines mark
$\alpha=0$ (Hartley limit) and $\alpha=1$ (Shannon entropy). The Meditation
curve is shown with increased line weight.}
\label{fig:renyi}
\end{figure}

\begin{figure}[H]
\centering
\includegraphics[width=0.60\linewidth]{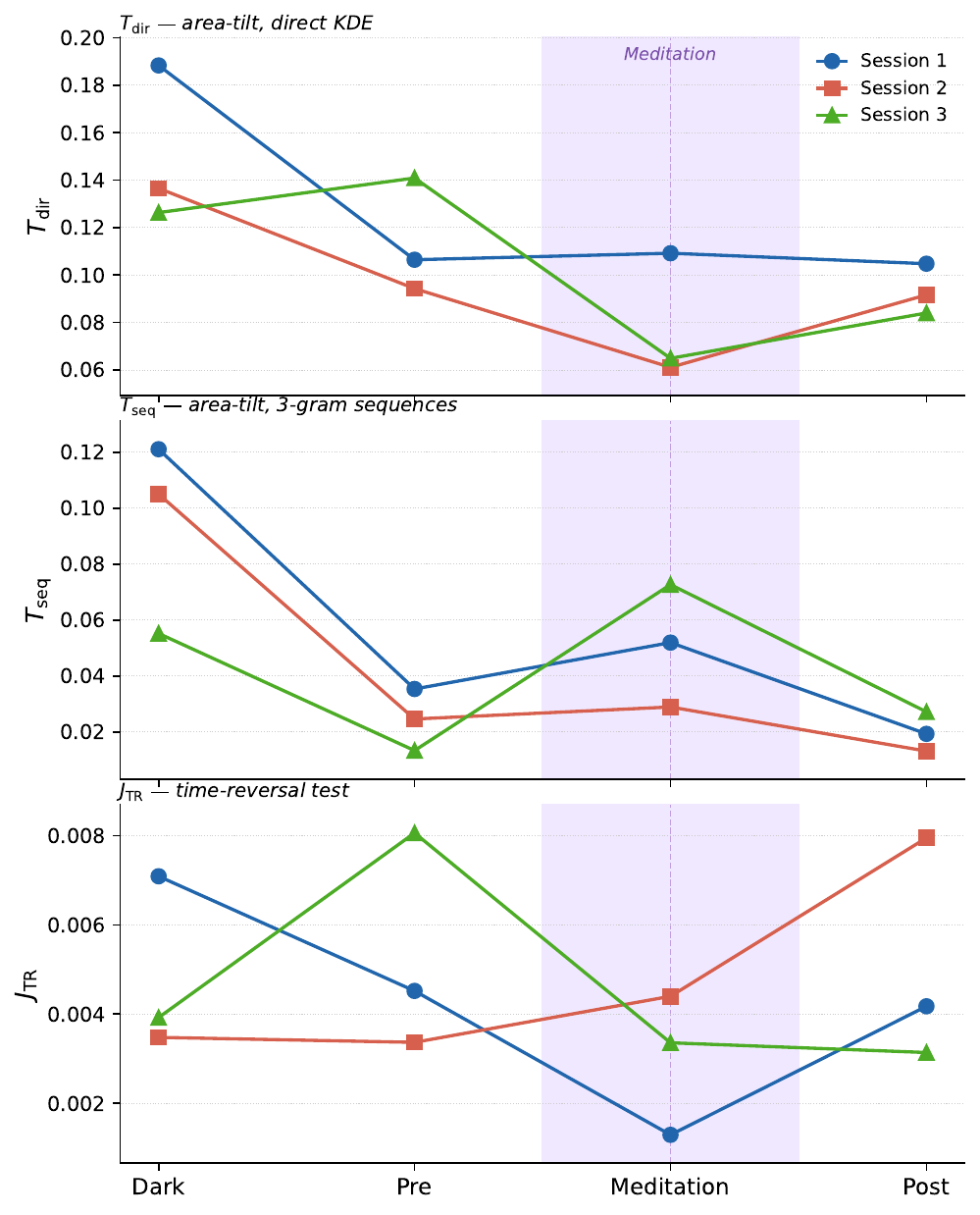}
\caption{R\'{e}nyi entropy scalar indices as a function of experimental phase
for all three sessions (blue circles: Session~1; red squares: Session~2;
green triangles: Session~3). Top: area-tilt $T_{\rm dir}$ (direct KDE
method). Middle: area-tilt $T_{\rm seq}$ (3-gram sequence method). Bottom:
Time Reversal index $J_{\rm TR}$. The shaded band marks the Meditation phase.
Lower $T$ indicates lighter-tailed distributions; lower $J_{\rm TR}$ indicates
dynamics closer to time-reversibility.}
\label{fig:renyi_indices}
\end{figure}

Figure~\ref{fig:renyi} shows the full $\tilde{H}_\alpha(\alpha)$ curves for
all three sessions and four phases using both methods.
The Dark phase is visually well separated from the active phases across the
entire $\alpha$ range in all panels, confirming that the instrumental
background has a qualitatively different distributional structure from the
biological signal. Among the active phases, the direct-method curves nearly
overlap, while the sequence-method curves display a modest but visible
stratification at $\alpha < 0$, reflecting session-dependent differences in
the short-range temporal organisation of the emission.
The Dark phase yields the highest $T_{\rm dir}$ in all three sessions
($0.126$--$0.188$), well separated from the active phase range
($0.061$--$0.141$). This separation reflects the Poissonian character of
the instrumental background at a count rate roughly half that of the
biological signal, and provides an independent confirmation of the
Dark/signal distinction already established by the mean count and Fano
factor. The $T_{\rm seq}$ index confirms the Dark separation in Sessions~1
and~2, while in Session~3 its Dark value ($0.055$) falls closer to the active
phase range, consistent with the anomalously elevated mean Dark count
($\mu = 5.36$) which modifies the sequential structure of the background
in that session.
Among the active phases, $T_{\rm dir}$ shows Meditation at the minimum in
Sessions~2 and~3 ($0.061$ and $0.065$ respectively), while in Session~1
the three active phases lie within $0.105$--$0.109$ and are
indistinguishable within fluctuations.
The directional consistency in two out of three sessions, Meditation
associated with the smallest tail-to-bulk asymmetry, is coherent with
the global reduction of super-Poissonian excess captured by the Fano factor
and Expected Shortfall in Section~\ref{sec:expdata}: all three independent
measures converge on the same picture of reduced emission burstiness during
the meditative state.
The scalar indices $T_{\rm dir}$, $T_{\rm seq}$, and $J_{\rm TR}$ extracted
from the curves of Figure~\ref{fig:renyi} are shown as a function of
experimental phase in Figure~\ref{fig:renyi_indices} for all three sessions.
%
%\FloatBarrier
Figure~\ref{fig:renyi_indices} collects the scalar observables extracted
from the R\'{e}nyi analysis and displays their evolution across the four
experimental phases in all three sessions. While the full
$\tilde{H}_\alpha(\alpha)$ curves of Figure~\ref{fig:renyi} provide a
detailed representation of the distributional structure, the indices shown
here allow a more compact comparison of the phase-dependent changes in
tail behaviour, sequential organisation, and temporal irreversibility of
the emission dynamics.
%\clearpage
\FloatBarrier
The Time Reversal index $J_{\rm TR}$ provides the sharpest individual result
in Session~1, where Meditation yields $J_{\rm TR} = 0.00129$, approximately
$3.5$ times lower than Pre ($0.00452$) and 3.2 times lower than Post ($0.00417$).
This reduction indicates that the short-range sequential dynamics of the
emission process approach time-reversibility during Sama Vritti breathing,
consistent with a transition toward a more regular, less intermittent regime.
Sessions~2 and~3 show a more complex pattern: in Session~2 the Post phase
reaches the highest $J_{\rm TR}$ while Pre and Meditation are comparable;
in Session~3 the Pre phase shows an anomalously high value ($0.00806$) while
Meditation and Post are both low ($0.00336$ and $0.00314$).
The inter-session variability of $J_{\rm TR}$ reflects its higher sensitivity
to transient non-stationarities relative to amplitude-based indices, and
likely also genuine day-to-day variation in the depth of the meditative state.
\vspace{12pt}
\paragraph{Comparative analysis of R\'{e}nyi indices and 
null-model validation.}
\label{sec:renyi_null}

\begin{figure}[htbp]
\centering
\includegraphics[width=0.82\linewidth]{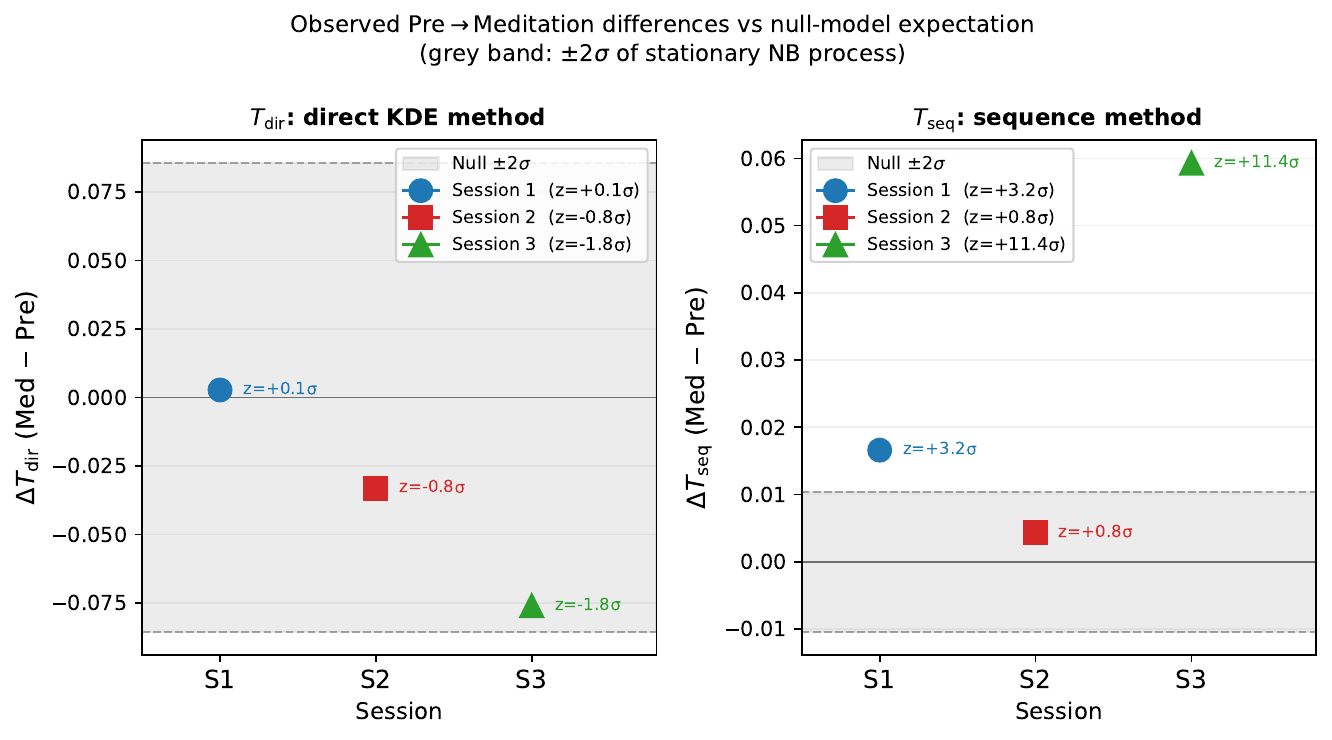}
\caption{Observed Pre$\to$Meditation differences $\Delta T_{\rm dir}$
(left) and $\Delta T_{\rm seq}$ (right) for the three experimental
sessions (blue circles: Session~1; red squares: Session~2; green
triangles: Session~3). The grey band marks the $\pm 2\sigma$ region
of the null model (stationary negative-binomial process with the same
$\mu$ and $F$ as the Pre phase; see Appendix). $z$-scores of the
observed differences relative to the null distribution are annotated
for each session. $\Delta T_{\rm dir}$ is consistently negative
(reduced tail-to-bulk asymmetry in the marginal distribution), while
$\Delta T_{\rm seq}$ is consistently and significantly positive
(increased sequential pattern structure), reflecting two distinct and
complementary effects of the meditative state on the emission dynamics.}
\label{fig:delta_renyi}
\end{figure}

The scalar indices extracted from the R\'{e}nyi analysis admit a more
direct statistical interpretation when the observed Pre$\to$Meditation
differences are expressed as absolute changes
$\Delta T = T_{\rm Med} - T_{\rm Pre}$ and compared against the
fluctuations expected from a stationary process. Figure~\ref{fig:delta_renyi}
shows $\Delta T_{\rm dir}$ and $\Delta T_{\rm seq}$ for all three sessions,
together with the $\pm 2\sigma$ band of the null model (a stationary
negative-binomial process with the same $\mu$ and $F$ as the Pre phase;
see Appendix for full details).

Two complementary patterns emerge from Figure~7. For $T_{\text{dir}}$ (left panel),
the Pre$\to$Meditation differences have z-scores of $+0.10$, $-0.80$, and $-1.82$
in Sessions~1, 2, and~3 respectively. Applying Stouffer's method across $k = 3$ sessions:
\begin{equation}
z_{\text{comb}}^{\text{dir}} = \frac{+0.10 - 0.80 - 1.82}{\sqrt{k}} = 
\frac{-2.52}{\sqrt{3}} \approx -1.45.
\end{equation}
Under Stouffer's method $z_{\text{comb}}^{\text{dir}}$ follows a standard normal
distribution; the observed value $z_{\text{comb}}^{\text{dir}} \approx -1.45$ therefore
gives $p = P(Z < -1.45) \approx 0.075$ (one-tailed), where $Z \sim \mathcal{N}(0,1)$,
a result that approaches but does not reach conventional significance, indicating
a consistent but weak reduction of tail-to-bulk asymmetry in the marginal amplitude
distribution during Meditation.

The picture for $T_{\rm seq}$ (right panel) is strikingly different.
Rather than decreasing, $T_{\rm seq}$ \emph{increases} from Pre to
Meditation in all three sessions, with $z$-scores of $+3.17$, $+0.81$,
and $+11.4$ in Sessions~1, 2, and~3 respectively. Sessions~1 and~3
lie far outside the null band, establishing a highly significant and
reproducible increase in sequential pattern structure during the
meditative phase. The index $T_{\rm seq}$ measures the tail-to-bulk
asymmetry of the R\'{e}nyi entropy computed on the distribution of
3-gram sequential patterns: a higher value indicates that certain
patterns are much more probable than others, i.e., the symbolic
dynamics of the emission become more concentrated on specific sequential
motifs. We interpret this as the signature of the Sama Vritti breathing
rhythm itself: the box-breathing cycle has a fixed period of 16~s
($4 \times 4$~s phases), which at 0.5~s resolution creates a structured,
repeating modulation of the biophoton count series. This periodic
modulation concentrates the 3-gram distribution on the patterns
associated with the breathing cycle, driving $T_{\rm seq}$ upward.

The opposite signs of $\Delta T_{\rm dir}$ and $\Delta T_{\rm seq}$
are not contradictory: they reflect two distinct physiological mechanisms
acting simultaneously during Sama Vritti meditation.
The decrease of $T_{\rm dir}$ is consistent with the reduction of
metabolic burstiness captured by the Fano factor and Expected Shortfall:
individual high-count events become less frequent and less intense,
making the marginal amplitude distribution lighter-tailed.
The increase of $T_{\rm seq}$, by contrast, reflects the imposition of
a structured respiratory rhythm on the emission dynamics: the periodic
breathing cycle creates repeating sequential patterns that are absent
during the unstructured Pre phase. These two effects operate on
different aspects of the signal --- its marginal amplitude statistics
versus its short-range temporal organisation --- and are therefore
simultaneously consistent. Taken together, they suggest that the
meditative state modulates biophoton emission on two levels: a
metabolic level, through the reduction of sympathetic-driven oxidative
bursts, and a rhythmic level, through the entrainment of emission
dynamics to the respiratory cycle.
This interpretation is further supported by the Time Reversal index
$J_{\rm TR}$, which decreases in Sessions~1 and~3 ($z = -1.13$ and
$-1.62$ respectively against the null model), indicating that the
short-range sequential dynamics approach time-reversibility during
Meditation --- a signature consistent with a system driven toward a more
regular, less dissipative regime by the periodic breathing constraint.
\vspace{12pt}
%\clearpage
\nopagebreak

\section{Discussion and Conclusions}
\label{sec:discussion}
The four independent analyses presented in this paper converge on a
coherent picture of how Sama Vritti meditation modifies the dynamics of
biophoton emission from the human palm, while exhibiting a degree of
session-to-session variability that reflects the intrinsic biological
complexity of a single-subject proof-of-concept study.

The most robust result, holding without exception across all three sessions 
and all analytical methods, is a systematic reduction of emission intermittency 
during the meditative phase relative to the pre-meditation baseline. The Fano 
factor $F$ and skewness $\gamma_1$ show a consistent Pre$\to$Meditation 
reduction in every session, reaching their minimum at Meditation in Sessions~2 
and 3, and continuing to decrease monotonically into Post in Session~1; the 
right-tail Expected Shortfall $\mathrm{ES}_\mathrm{dx}$ decreases consistently 
from Pre to Meditation, indicating that not only the frequency but also the 
average intensity of extreme high-count events is reduced; the multiscale Fano 
analysis confirms that this suppression grows systematically across all 
averaging scales $\tau$. Treating the three sessions as independent replications, 
Fisher's combined test over the four distributional indices ($F$, $\gamma_1$, 
$\mathrm{ES}_\mathrm{dx}$, $\mathrm{ES}_\mathrm{sx}$) yields $p \approx 0.034$, 
establishing that the Pre$\to$Meditation reduction is statistically significant 
at the level of the full distributional characterisation, while the directional 
consistency across all sessions and all indices remains the primary 
assumption-free evidence.

A second, complementary layer of evidence comes from the complexity analyses. 
The Allan deviation exponent $\alpha_A$ is consistently least negative in the 
Pre phase, indicating stronger temporal correlations; the Pre$\to$Meditation 
difference is statistically significant in Sessions~1 and 3 ($3.0\sigma$ and 
$2.1\sigma$ respectively), while in Session~2 the difference is negligible 
($0.2\sigma$), consistent with the regularisation in that session operating 
primarily at shorter time scales, where the Fano factor shows its largest 
suppression across all three sessions.

The stripe-filtered DEA yields $\delta > 0.5$ in the majority of biological 
phases across all three sessions, placing the emission dynamics in the 
non-ergodic, non-stationary renewal regime ($\mu \approx 2.5$--$2.9$) 
consistent with previous results on cell cultures and germinating seeds from 
our group (Benfatto et al., 2021; De Paolis et al., 2024, 2026). The 
Pre$\to$Meditation change of $\delta$ shows a consistent directional pattern: 
Session~2 yields a significant decrease ($\Delta\delta = -0.132$, ${\sim}2.6\sigma$); 
Session~1 shows a decrease in the same direction but within the combined 
measurement uncertainties ($\Delta\delta = -0.049$, ${\sim}0.9\sigma$); 
Session~3 shows a negligible increase ($\Delta\delta = +0.030$, ${\sim}0.7\sigma$) 
that is indistinguishable from zero and therefore non-contradictory. Combining 
the three $z$-scores via Stouffer's method yields $z_\mathrm{comb} \approx -1.60$ 
($p \approx 0.055$, one-tailed), approaching conventional significance: the 
overall directional evidence consistently points toward a reduction of the 
renewal correlation strength during Sama Vritti breathing.

The R\'{e}nyi analysis adds a further dimension. The area-tilt 
$T_\mathrm{dir}$ is minimised in Meditation in Sessions~2 and 3; in 
Session~1 the three active phases are indistinguishable within fluctuations. 
Combining the three sessions via Stouffer's method yields $p \approx 0.075$ 
(one-tailed), indicating a consistent but moderate reduction of tail-to-bulk 
asymmetry in the marginal amplitude distribution. The Time Reversal index 
$J_\mathrm{TR}$ shows reductions in Sessions~1 and 3 ($z = -1.13$ and 
$-1.62$ respectively relative to the null model), indicating an approach to 
time-reversibility during Sama Vritti breathing; in Session~2, Pre and 
Meditation are comparable while the Post phase reaches the highest value, 
reflecting the higher sensitivity of $J_\mathrm{TR}$ to transient 
non-stationarities. The sequence-method area-tilt $T_\mathrm{seq}$ reveals 
a complementary and highly systematic pattern: $T_\mathrm{seq}$ increases 
from Pre to Meditation in all three sessions, with $z$-scores of $+3.17$, 
$+0.81$, and $+11.4$ relative to the stationary null model. This result, 
significant in Sessions~1 and 3, indicates that the distribution of 
sequential 3-gram patterns becomes more concentrated during Meditation~--- 
a signature consistent with the entrainment of the emission dynamics to the 
regular 16~s rhythm of the Sama Vritti breathing cycle.

The analytical methods employed here do not all reach the same level of
significance in every session, but this reflects their complementary
sensitivities rather than genuine contradictions: within each individual
session, no index points in a direction inconsistent with the others.
In Session~2, the Allan deviation Pre$\to$Meditation difference is negligible
($0.2\sigma$) while the Fano factor shows the largest suppression of all three
sessions; this reflects the fact that the two indices probe different aspects
of the same process, the Allan deviation is sensitive to
consecutive-interval correlations at scales $\tau$--$2\tau$, while the Fano
factor integrates the variance excess over all scales.
The coincidence of significant Fano suppression with a flat Allan deviation
in Session~2 suggests that in that session the regularisation induced by
meditation operates primarily at shorter time scales, captured by $F$ but not
by $\alpha_A$.

Similarly, the R\'{e}nyi $T_{\rm dir}$ is insensitive to Session~1 because
the active-phase distributions are nearly indistinguishable in the marginal
amplitude domain, while the Time Reversal test $J_{\rm TR}$ reveals a strong
effect in the sequential dynamics of the same session.
The full battery of methods is therefore not redundant but necessary: each
method resolves aspects of the meditative state that the others do not.
To verify that the Pre$\to$Meditation differences in $\alpha_A$ and the
R\'{e}nyi indices are not attributable to statistical fluctuations between
consecutive windows of a stationary process, we performed a null-model
validation based on a stationary negative-binomial process with the same
marginal statistics as the Pre phase; the results, reported in the Appendix,
confirm that the observed differences exceed the expected null fluctuations
and are consistent with a genuine physiological transition.

The observed session-to-session variability in the magnitude and
pattern of the effects is expected and biologically meaningful even within a
single subject.
The depth of the meditative state, the degree of autonomic relaxation, and
the general physiological arousal level of the subject vary from day to day in
ways that are not controlled parameters in a single-session protocol.
This variability motivates repeated sessions in any physiological study
and ultimately requires adequate sample sizes for population-level inference.
Nevertheless within each session the different analytical methods consistently
identify measurable dynamical differences between the Pre, Meditation, and
Post phases, supporting the conclusion that the biophoton emission process
is sensitive to controlled physiological modulation.
Human UPE measurements are intrinsically challenging because of the weakness
of the signal, the short accessible acquisition times, and the unavoidable
biological variability associated with meditative practice.
The strategy adopted here, combining multiple complementary methods on
repeated sessions, is therefore intended primarily to identify robust
observables and to determine which analytical tools are most sensitive to
physiological state transitions in realistic experimental conditions.

\nopagebreak
The transition observed at the Pre$\to$Meditation boundary has a natural
biological interpretation, though we emphasise that what follows is
necessarily conjectural in the absence of direct biochemical measurements.
Biophoton emission from the palm originates primarily from the radiative
decay of electronically excited molecular intermediates generated during
oxidative metabolism , reactive oxygen species, lipid peroxidation
products, and excited carbonyl compounds \citep{Popp1,Wijk,CifraPospisil2014}.
Heavy-tailed, super-Poissonian emission statistics with strong temporal
correlations (high $F$, large $\mathrm{ES}_{\rm dx}$, $\delta$ well above 0.5)
are therefore the signature of an intermittent, episodic oxidative activity
, bursts of reactive species followed by quiescent intervals , of the
kind associated with high sympathetic tone.
The systematic transition toward lighter tails, weaker temporal correlations
(scaling approaching Poisson), and reduced temporal irreversibility observed
in the Pre$\to$Meditation comparison is consistent with a shift of the
oxidative dynamics in the palm tissue toward a more regular, less intermittent
regime.
We conjecture that this reflects the well-documented parasympathetic
activation induced by slow, patterned breathing at $\sim$3.75 cycles per
minute: reduced sympathetic drive would suppress episodic, catecholamine-mediated
oxidative bursts, yielding a more uniform ROS production rate and consequently
a biophoton emission process closer to a Poissonian baseline.

The simultaneous increase of $T_{\rm seq}$ adds a second, distinct layer
to this picture: beyond the metabolic regularisation captured by $F$,
$\mathrm{ES}_{\rm dx}$, $T_{\rm dir}$, $\alpha_A$, and $\delta$, the
periodic structure of the Sama Vritti breathing rhythm is directly imprinted
on the sequential organisation of the emission, driving the 3-gram pattern
distribution toward a more concentrated, rhythmically structured state.
These two effects --- metabolic quietening and respiratory entrainment ---
operate on different aspects of the signal and are mutually consistent:
the former reduces sympathetic-driven oxidative bursts, the latter imposes
a deterministic periodic modulation on the resulting emission stream.
The approach to time-reversibility ($J_{\rm TR} \to 0$) is particularly
suggestive: in a thermodynamic context, lower entropy production per unit
time is the hallmark of a system operating closer to its minimum dissipation
state, consistent with the known reduction in oxygen consumption and
metabolic rate reported during deep meditation.
\nopagebreak[4]
From a broader physiological perspective, the results of this study suggest
that biophoton emission from the human hand provides a sensitive and
informationally rich probe of the metabolic changes associated with
structured breathing and meditative practice.

Several EEG studies have documented that practices of this type, including box
breathing and related protocols, induce characteristic delta and theta power
increases, increased theta and high-beta connectivity, and phase-amplitude
coupling between these bands in prefrontal and default-mode-network regions
\citep{Zaccaro2022}. Of particular relevance, Zaccaro et al. \ employed precisely the same
Sama Vritti Pranayama protocol and observed these EEG aftereffects in a cohort
of 12 experienced meditators measured after the breathing session.
The convergence with the present findings, obtained in a single non-expert
subject and during the session itself rather than in post-session
recovery, supports the interpretation that UPE provides a real-time
metabolic readout of the same autonomic state transition, detectable across
independent physiological channels and independently of the practitioner's
level of meditative expertise.

At the same time, the HRV analysis has shown
a transition toward more coherent and less intermittent cardiac dynamics,
quantifiable via DEA: specifically, Tuladhar et al.\ demonstrated that
meditation shifts the renewal index $\mu$ of the heartbeat inter-event
distribution from values near $\mu \approx 2$ (ideal $1/f$ regime) toward
$\mu \approx 3$ (Gaussian basin of attraction), corresponding to a decrease
of the DEA scaling exponent $\delta$ toward the Poisson reference --- 
precisely the same directional transition observed here in the biophoton
channel \citep{Tuladhar2018,Jelinek2021}.

The present results indicate that analogous transitions are detectable in
the biophoton channel, but through a fundamentally different physiological
pathway: whereas EEG reflects electrical neural dynamics and ECG reflects
cardiac autonomic modulation, UPE is directly linked to cellular oxidative
metabolism and the radiative decay of electronically excited reactive oxygen
species \citep{Popp1,Wijk,CifraPospisil2014}.

This interpretation is conceptually consistent with the recent multimodal
study of Dyer et al.\ \citep{dyer2026changes}, in which 23 subjects were
simultaneously monitored with EEG, HRV, skin conductance, infrared, and UPE
sensors during a loving kindness meditation followed by a breathwork exercise.
Notably, it is during the breathwork phase --- structured, patterned breathing
--- that left hand UPE shows a near-significant decreasing trend ($p = 0.057$),
while the meditation phase produces a significant increase in nasal infrared
emission. Since our Sama Vritti protocol is itself a structured breathing
practice, the natural comparison is with Dyer's breathwork phase rather than
their meditation phase: both results point to a modulation of UPE by
rhythmic respiratory control, consistent with the respiratory entrainment
signature identified here via $T_{\rm seq}$. The present work extends this
perspective by targeting not the mean emission level but the temporal
organisation and dynamical complexity of the UPE signal, providing a
complementary and more detailed characterisation of the same phenomenon.

Taken together, the convergence of these different physiological channels
toward a picture of increased regularity and reduced intermittency during
meditation suggests a coordinated state transition that spans multiple
physiological levels simultaneously --- neural, autonomic, and metabolic ---
a conclusion that neither channel alone could support.
Simultaneous multi-channel recordings combining EEG, ECG, and UPE would allow
this hypothesis to be tested directly, providing a quantitative cross-modal
characterisation of the meditative state that goes substantially beyond what
is currently available in the literature.

\section{Future Perspectives}
\label{sec:future}
The present work constitutes a proof-of-concept study on a single subject
over three sessions: it establishes the viability of the measurement
protocol, identifies the most sensitive analytical indices, and provides
the methodological foundation for a systematic multi-subject investigation.

The immediate next step is a careful determination of the required sample
size.
The analytical methods employed here show markedly different sensitivities
to the Pre$\to$Meditation transition: the Fano factor produces a consistent
and significant effect in all three sessions, suggesting that a relatively
small sample of $N \sim 5$--$10$ subjects would suffice for a
population-level detection; the Allan deviation and R\'{e}nyi indices are
moderately less consistent across sessions and point to a comparable or
slightly larger requirement; 
the stripe-filtered DEA exponent $\delta$ is the most demanding, as the
Pre$\to$Meditation decrease is clearly significant in one session only
(Session~2, ${\sim}2.6\sigma$) and approaches overall significance via
Stouffer combination ($p \approx 0.055$, one-tailed), placing its
sample-size requirement at $N \sim 15$ or above.
A conservative but experimentally realistic target of $N = 15$--$20$
subjects would therefore satisfy the power requirements of the full
analytical battery, and constitutes the primary goal of the next phase
of this investigation.

A second priority is the extension of the protocol to experienced
practitioners of different meditative traditions.
The subject of the present study, while acquainted with meditation,
is not an advanced practitioner, which motivated the choice of Sama Vritti
box breathing , a simple, highly structured protocol well-suited to
na\"{i}ve participants.
Expert meditators from traditions such as Kundalini Yoga, Zen, and Tibetan
Buddhism would offer several complementary advantages: longer and more
reproducible meditative states, access to deeper levels of absorption, and
the possibility of extended sessions exceeding the 15-minute phases used
here, providing longer time series and a wider scaling range for the
complexity analyses. In particular, we have planned to make extended measurements on expert practitioners of the Rinzai Zen school, belonging to the lineage of Yamada Mumon Roshi, introduced in Europe by Engaku Taino (Luigi Mario) and currently directed by Master Mario Nanmon Fatibene. In this way we have access to a well-established tradition in which prolonged sesshin practice allows experienced practitioners to sustain stable meditative states over extended periods, providing longer and more reproducible time series for complexity analysis.

Comparing results across traditions would allow investigation of whether the
biophoton signatures identified in this work reflect a common physiological
response to the meditative state independent of technique, or whether
tradition-specific features , such as the presence or absence of breathing
control, visualisation, or mantra , leave distinct signatures in the
emission dynamics.
Finally, we are currently developing a dedicated portable measurement system
designed specifically for this class of experiments.
The new setup is conceived as a wearable device , essentially an
instrumented glove , accommodating photon-counting detectors for both
hands simultaneously.
Bilateral measurement would allow real-time comparison of left and right
palm emission, providing access to lateralisation effects that are inaccessible
with the single-hand configuration used here, and would double the statistical
information available per session.
The portable design would furthermore enable measurements in naturalistic
settings outside the laboratory, reducing the logistic constraints of the
current protocol. 

An additional natural extension of the present work is the simultaneous
acquisition of biophoton emission together with physiological signals such
as ECG and EEG\@.
Among these possibilities, ECG appears particularly advantageous as a first
step because it is substantially less invasive, more stable over long
recording sessions, and less sensitive to motion artefacts than EEG\@.
In the context of meditative protocols, EEG recordings are often perturbed
by eye blinks, facial muscle activity, tongue motion, and small postural
adjustments, whereas wearable ECG systems can provide continuous and reliable
recordings with comparatively simple experimental logistics.
Modern portable ECG devices , including lightweight chest-band or
patch-based systems , are now capable of producing accurate RR-interval
series with precise time synchronisation, making them especially suitable
for simultaneous acquisition with UPE measurements.
From a methodological perspective, ECG is also naturally connected to the
complexity-analysis framework adopted in the present study: DEA, Fano
scaling, renewal statistics, and related approaches have already proven
informative in heart-rate variability analysis during meditation and
autonomic modulation.
Simultaneous UPE--ECG recordings would therefore make it possible to
investigate whether the dynamical transitions observed in the biophoton
channel are temporally correlated with changes in cardiac dynamics, and
would open the possibility of applying directional information-theoretic
tools such as Transfer Entropy to quantify cross-modal interactions between
metabolic and autonomic processes during meditation.

This direction is further motivated by previous studies on meditation and
heart-rate variability based on Diffusion Entropy Analysis and crucial-event
dynamics, particularly the work of Tuladhar and Jelinek \citep{Tuladhar2018,Jelinek2021}\ and related
investigations involving Gemignani \citep{Allegrini3}\ and collaborators.
Those studies demonstrated that meditative practices induce measurable
transitions in cardiac complexity and autonomic organisation.
Simultaneous ECG and UPE recordings would therefore provide a natural bridge
between the autonomic-cardiac domain already explored through DEA and the
metabolic-photonic channel investigated in the present work.
In this perspective, the key question would no longer be limited to whether
meditation modifies biophoton emission, but whether the transitions observed
in cardiac and metabolic complexity emerge coherently across physiological
levels and possibly exhibit directional coupling , a hypothesis that could
be explored through Transfer Entropy and related causality measures.

\vspace{6pt}
\section*{Appendix: Null-Model Validation for Allan 
Deviation and R\'{e}nyi Entropy}
\label{app:null_test}

The complexity indices reported in Sections~\ref{sec:allan}
and~\ref{sec:renyi} --- the Allan deviation exponent $\alpha_A$ and the
R\'{e}nyi scalar indices $T_{\rm dir}$, $T_{\rm seq}$, and $J_{\rm TR}$ ---
are estimated from time windows of approximately 1800 bins (15 minutes at
0.5~s resolution). A natural question is whether the Pre$\to$Meditation
differences observed in these indices could arise from statistical
fluctuations between consecutive windows of a stationary process, rather
than from a genuine physiological transition. To address this quantitatively,
we construct a stationary null model and compare the observed differences
against the distribution of differences expected under stationarity.

The null model is a stationary negative-binomial (NB) process. This choice
is motivated by the fact that the NB distribution reproduces the
super-Poissonian marginal statistics of the biological signal --- mean $\mu$
and Fano factor $F > 1$ --- while containing no temporal structure beyond
shot noise. A NB process with mean $\mu$ and Fano factor $F$ is parameterised
by $n = \mu/(F-1)$ and $p = 1/F$, giving $\mathbb{E}[X] = \mu$ and
$\mathrm{Var}[X]/\mathbb{E}[X] = F$ by construction. For each session,
$\mu$ and $F$ are set to the values measured in the Pre phase
(Table~\ref{tab:cross_stats}), so that the null model has the same marginal
distribution as the baseline from which the meditative transition is measured.

For each of $N_{\rm sim} = 500$ realisations, a stationary NB series of
$2 \times 1800$ bins is generated and split into two consecutive windows of
1800 bins each --- mimicking the Pre and Meditation phases in terms of length.
The indices $\alpha_A$, $T_{\rm dir}$, $T_{\rm seq}$, and $J_{\rm TR}$ are
computed for each window using the same algorithms employed on the experimental
data: fractional Allan deviation with power-law fit over
$\tau = 0.5$--$128$~s; R\'{e}nyi KDE and 3-gram sequence methods with
$\alpha \in [-8, 6]$, $L = 3$, $K = 4$. The difference between the second
and first window gives one null realisation of $\Delta\mathrm{index}$. After
$N_{\rm sim}$ realisations, the null distribution of each $\Delta\mathrm{index}$
has mean $\mu_0 \approx 0$ and standard deviation $\sigma_0$. The $z$-score
of the observed Pre$\to$Meditation difference is then:
\begin{equation}
    z = \frac{\Delta\mathrm{index}_{\rm obs} - \mu_0}{\sigma_0}.
    \label{eq:zscore_null}
\end{equation}

\begin{figure}[htbp]
\centering
\includegraphics[width=0.82\linewidth]{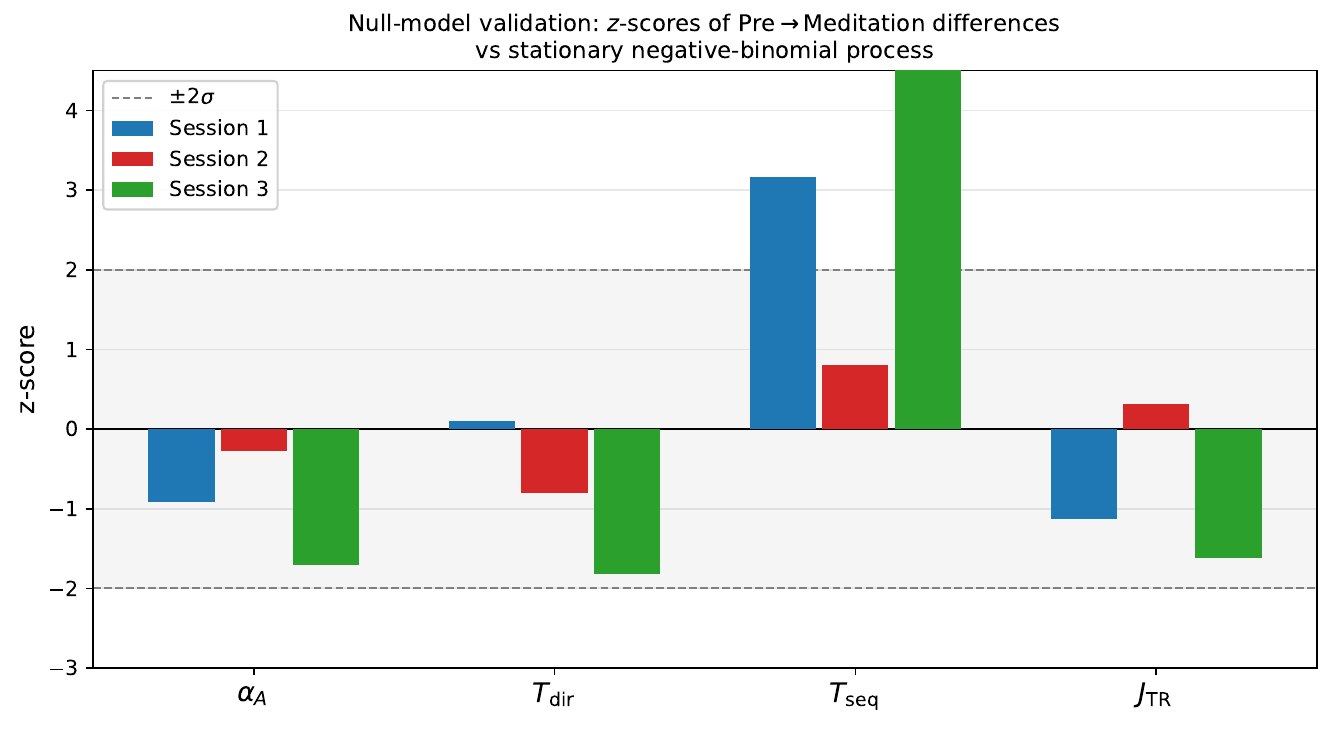}
\caption{Null-model validation: $z$-scores of the observed
Pre$\to$Meditation differences in $\alpha_A$, $T_{\rm dir}$, $T_{\rm seq}$,
and $J_{\rm TR}$ (blue: Session~1; red: Session~2; green: Session~3),
computed relative to the null distribution obtained from $N_{\rm sim} = 500$
realisations of a stationary negative-binomial process with the same $\mu$
and $F$ as the Pre phase. The grey band and dashed lines mark the $\pm 2\sigma$
region of the null model. The $z$-score for $T_{\rm seq}$ in Session~3
($z = +11.4$) lies far outside the plotted range.}
\label{fig:null_test}
\end{figure}

Figure~\ref{fig:null_test} shows the $z$-scores for all four indices and all
three sessions. All null distributions are centred near zero
($|\mu_0| \ll \sigma_0$), confirming that a stationary NB process does not
produce systematic Pre$\to$Meditation differences in any index: any difference
observed in the experimental data must therefore originate from a genuine
change in the dynamical organisation of the emission, not from sampling
fluctuations of a stationary process.

Examining Figure~\ref{fig:null_test} in detail: the Allan deviation exponent
$\alpha_A$ shows negative $z$-scores in all three sessions, indicating that
$\alpha_A$ is consistently more negative during Meditation than Pre,
while the null model predicts no preferred direction; the probability of
observing three concordant negative values by chance is $(1/2)^3 = 0.125$.
The index $T_{\rm dir}$ shows reductions in Sessions~2 and~3 ($z = -0.80$
and $-1.82$ respectively), consistent with the reduced distributional
burstiness reported in Section~\ref{sec:renyi}. The Time Reversal index
$J_{\rm TR}$ shows reductions in Sessions~1 and~3 ($z = -1.13$ and $-1.62$
respectively), consistent with the approach to time-reversibility during
Sama Vritti breathing. Most strikingly, $T_{\rm seq}$ increases significantly
above the null in Sessions~1 and~3 ($z = +3.17$ and $+11.4$ respectively),
pointing to a genuine increase in sequential pattern structure during
Meditation that is discussed in Section~\ref{sec:renyi} and
Section~\ref{sec:discussion}.

While none of the four indices individually produces a $z$-score that would
be considered conclusive in isolation, the convergence of all four indices
across three independent sessions --- with null distributions centred at zero
in every case --- makes the overall picture substantially more robust than
any single index could establish alone.

This null-model validation is fully consistent with, and complementary to,
the stripe-filtered DEA result of Section~\ref{sec:dea}. The DEA already
establishes that the biological signal cannot be described as a stationary
Poissonian process: the Dark phase systematically yields $\delta \approx 0.5$
while all biological phases yield $\delta > 0.5$ in all three sessions, a
separation that the stripe filter makes immune to Poissonian noise
contributions by construction \citep{Allegrini1, BenfattoArXiv2025}.
The present null test extends this conclusion to $\alpha_A$ and the R\'{e}nyi
indices, showing that the Pre$\to$Meditation differences in these quantities
are also inconsistent with fluctuations expected from a stationary process
with the same marginal statistics as the data. Taken together, the two
results provide converging and independent evidence that the observed dynamical
transitions are genuine physiological effects.

\vspace{3pt}
\section*{Acknowledgements}
The authors gratefully acknowledge I. Davoli and R. Francini for their indispensable contribution to the acquisition of the experimental data. We warmly thank Master Mario Nanmon Fatibene, Abbot of the Rinzai Zen school of the Yamada Mumon lineage, for his kind availability and openness to participating in future experimental sessions. The authors owe special gratitude to P. Grigolini of the University of North Texas (UNT), whose invaluable discussions, unwavering encouragement, and pioneering vision have been the true driving force behind this work. Without him this work would never have seen the light.
\vspace{12pt}
\bibliographystyle{unsrt}
\bibliography{Bib}
\end{document}